\documentclass[twocolumn]{aastex63}

\usepackage{txfonts}
\usepackage{bm}

\newcommand{\pd}{\partial}
\newcommand{\au}{{\rm au}} 
\newcommand{\kB}{k_{\rm B}}
\newcommand{\paren}[1]{\left(#1\right)} 
\newcommand{\subsc}[2]{#1_{{\rm #2}}}
\newcommand{\eqref}[1]{(\ref{#1})}

\begin{document}

\title{Effects of Dust Evolution on the Vertical Shear Instability in the Outer Regions of Protoplanetary Disks}

\author[0000-0002-9660-8947]{Yuya Fukuhara}
\author[0000-0002-1886-0880]{Satoshi Okuzumi}
\author[0000-0001-8524-6939]{Tomohiro Ono}

\affiliation{Department of Earth and Planetary Sciences, 
Tokyo Institute of Technology, Meguro, Tokyo 152-8551, Japan}

\correspondingauthor{Yuya Fukuhara}
\email{fukuhara.y.ab@m.titech.ac.jp}

\shortauthors{Fukuhara et al.}

\submitjournal{ApJ}

\begin{abstract}
    The vertical shear instability (VSI) is a hydrodynamical instability that requires rapid gas cooling and has been suggested to operate in outer regions of protoplanetary disks.
    The VSI drives turbulence with strong vertical motions, which could regulate the dust growth and settling. 
    However, dust growth and settling can regulate the VSI because dust depletion makes gas cooling inefficient in outer disk regions that are optically thin to their own thermal emission.
    In this study, we quantify this potentially stabilizing effects of dust evolution on the VSI based on the linear analysis. 
    We construct a model for calculating the cooling timescale, taking into account dust growth beyond micron sizes and size-dependent settling.
    Combining the model with the linear stability analysis, we map the region where the VSI operates, which we call the VSI zone, and estimate the maximum growth rate at each radial position. 
    We find that dust growth as well as settling makes the VSI zone more confined around the midplane. 
    This causes a decrease in the growth rate because the vertical shear of the rotation velocity, which is the source of the instability, is weaker at lower altitude.
    In our default disk model with 0.01 solar masses, dust growth from $10~\micron$ to $1{\rm ~mm}$ causes a decrease in the growth rate by a factor of more than 10.
    The suppression of VSI-driven turbulence by dust evolution may promote further dust evolution in the outer regions and also explain a high degree of dust settling observed in the disk around HL Tau. 

\end{abstract}


\section{Introduction}\label{sec:intro}
    
    { Planet formation begins with the evolution of dust grains in protoplanetary disks into kilometer-sized planetesimals. This first stage is initially driven by the growth of dust grains through mutual sticking and condensation \citep{Chokshi:1993aa,Dominik:1997aa}. 
    Large dust particles settle to the midplane and may experience the streaming instability \citep{Youdin:2005aa,JohansenYoudin2007} and gravitational instabilities \citep{Goldreich:1973aa,Youdin:2011aa,Takahashi:2014wi,Tominaga:2018th,Tominaga:2019uu,Tominaga:2020wn,Pierens:2021ve}, which concentrate the dust particles in a runaway fashion and thereby form planetesimals \citep[e.g.,][]{Johansen:2009aa,Carrera:2015aa,Yang:2017aa}. The dust particles may also grow directly into planetesimals if the particles are sticky enough  \citep[e.g.,][]{Okuzumi+2012,Windmark:2012aa,Kataoka:2013aa}. 
    }
    
    { Dust evolution  in protoplanetary disks depends on gas disk turbulence in many ways.  Turbulence enhances the relative velocity of solid aggregates and may prevent them from sticking together through collisions \citep[e.g.,][]{Brauer:2008aa,Okuzumi:2012aa}.
    Turbulence may also inhibit dust settling toward the disk midplane and formation of a dense dust layer at the midplane \citep[e.g.,][]{Dubrulle+1995}. 
    Therefore, constraining the level of turbulence in real protoplanetary disks is essential for fully understanding how the dust in the disks evolves into planetesimals.
    }
    
    { Recent radio interferometric observations with the Atacama Large Millimeter-submillimeter Array have provided us detailed information of dust evolution and gas turbulence in the outer part of protoplanetary disks.
    The observations have provided us with ample evidence that massive and large disks commonly have rings and gaps of dust \citep[e.g.,][]{ALMA+2014,Andrews+2018,Long:2018aa,van-der-Marel:2019aa}. Although there are a number of potential mechanisms that provide such substructures \citep[for a review, see][]{Andrews:2020aa}, many of them assume that the dust particles comprising the rings have already grown to $0.1$--10 mm in size so that they can concentrate radially under the influence of gas drag.
    Furthermore, the well separated morphology of the dust rings in the disk around HL Tau \citep{ALMA+2014} indicates that the dust particles comprising the rings have already settled significantly, with a dust scale height being ten times smaller than the gas scale height \citep{Pinte:2016aa}. Assuming that the dust rings are indeed dominated by millimeter-sized particles, the high degree of settling also points to a low level of turbulence near the midplane.
    Molecular line emission observations suggest that turbulence in the upper layers of the outer disk regions is also weak \citep{Flaherty:2015aa,Flaherty:2017aa,Flaherty:2018aa,Flaherty:2020aa}.
    }
    
    { The absence of strong turbulence in the outer disk regions is consistent with the theoretical expectation that the magnetorotational instability \citep[MRI,][]{BalbusHawley1991} in the outer regions is suppressed by ambipolar diffusion \citep{Simon+2013a,Simon+2013b,Bai2015,Riols:2018aa}. 
    However, the suppression of the MRI is not enough to explain the absence of strong turbulence because purely hydrodynamic disk instabilities can also drive turbulence \citep[for a review, see][]{LyraUmurhan2019}. 
    Among them, the most robust one in outer regions of protoplanetry disks is the vertical shear instability  \citep[VSI,][]{UrpinBrandenburg1998,NelsonGresselUmurhan2013,LinYoudin2015}.
    The VSI is an instability caused by a vertical gradient in the gas rotation velocity together with a cooling timescale { much} shorter than the orbital timescale \citep{Urpin2003,NelsonGresselUmurhan2013,LinYoudin2015}.
    The requirement of rapid gas cooling tends to be met in outer disk regions where the optical depth is low \citep{Malygin+2017,PfeilKlahr2019}.
    Once the VSI operates, it produces turbulence with predominant vertical motion \citep[e.g.,][]{StollKley2014}, which would efficiently prevent vertical dust settling \citep{FlockNelson+2017,Flock:2020aa}. 
    }

    { The question then is what can suppress the VSI in outer regions of protoplanetary disks. One candidate is strong magnetic field as suggested by \citet{NelsonGresselUmurhan2013} and \citet{Cui:2020aa}. 
    In this paper, we explore the potential role of dust evolution, i.e., dust growth and settling, in suppressing the VSI. It is natural to expect that dust evolution should affect the VSI as it is the dust that is responsible for disk cooling. 
    \citet{Malygin+2017} already noted that a depletion of small grains slows down cooling in { disk regions that are optically thin to their own thermal emission}. This implies that both dust evolution growth and settling should lead to suppression of the VSI, although no quantitative assessment of the effects has been made so far.
    }

    { The goal of this study is to clarify the influences of dust growth and settling on the stability of outer (5--100 \au) protoplanetary disk regions against the VSI. 
    We calculate the thermal relaxation (cooling) timescale in the outer disk regions using a parametrized dust model in which the maximum particle size and dust vertical diffusion coefficient are given as free parameters.
    Using the two-dimensional maps of the thermal relaxation timescale together with linear stability analysis, we study how the location of the VSI-active region and the VSI growth rate in the region vary as dust grows and settles.
    }

    { This paper is organized as follows. 
    In Section \ref{sec:VSI}, we review the basic properties of the VSI, deriving the linear dispersion relation that gives the growth rate of the VSI at each location in a disk.
    We then describe our model in Section~\ref{sec:model}, present the main results in Section \ref{sec:results}, and discuss limitations and implications of our study in Section \ref{sec:discussion}.
    Section \ref{sec:conclusions} presents a summary.
    }
    
\section{The VSI}\label{sec:VSI}
    
    { In this section, we review the basic properties of the VSI and derive the dispersion relation that is used in the following section. The VSI is a type of the Goldreich--Schubert--Fricke instability \citep{GS67,Fricke:1968aa} known in the context of differentially rotating stars. The presence of a vertical gradient in gas angular velocity $\Omega$ is one of the necessary conditions for the VSI to operate. The radial and vertical force balances give
        \begin{equation}\label{eq:Omega}
            \Omega^2 = \frac{GM_{\ast}}{\paren{R^2+z^2}^{3/2}}+\frac{1}{R\rho_g}\frac{\partial P}{\partial R},
        \end{equation}
        \begin{equation}\label{eq:vertical_shear}
            \frac{\partial \paren{R\Omega}}{\partial z} = \frac{1}{2\Omega \rho_g^2}\paren{\frac{\partial P}{\partial z}\frac{\partial \rho_g}{\partial R}-\frac{\partial P}{\partial R}\frac{\partial \rho_g}{\partial z}},
        \end{equation}
    where $R$ is the cylindrical distance from the central star, $z$ is the height from the midplane, $M_{\ast}$ is the mass of the central star, $P$ is the gas pressure, $\rho_g$ is the gas density, and $G$ is the gravitational constant. The vertical shear $\partial(R\Omega)/\partial z$ is nonzero if the temperature gradient exists in the radial direction \citep[][see also Equation \eqref{eq:vertical_shear2} in Section \ref{subsec:disk}]{Urpin2003}. 
    }

    \subsection{The Thermal Criterion for Instability}\label{subsec:VSI_principle}
        { However, buoyant forces can stabilize the VSI when the entropy increases in the direction of decreasing gas pressure \citep{LinYoudin2015}. The Brunt--V\"{a}is\"{a}l\"{a} frequency $N_z$ is given by 
        \begin{equation}\label{eq:Nz}
            N_z^2 \equiv -\frac{1}{\rho_gC_P}\cdot\frac{\partial P}{\partial z}\cdot\frac{\partial s}{\partial z},
        \end{equation}
        where $C_P$ and $s$ are the specific heat at constant pressure and the specific entropy, respectively.
        {The specific entropy is given by $s = C_V\log{\paren{P/\rho_g^\gamma}}$, where $C_V$ is the specific heat at constant volume and $\gamma$ is the heat capacity ratio.}
        The vertical buoyancy is stabilizing if $N_z^2 > 0$. In protoplanetary disks, the pressure decreases with distance $z$ from the midplane. 
        In outer disk regions where the temperature is determined by stellar irradiation \citep[e.g.,][]{Chiang:1997aa}, the entropy increases with $z$, thus stabilizing the VSI.
        }

        { Therefore, the VSI requires fast thermal relaxation that reduces buoyancy  \citep{NelsonGresselUmurhan2013}. This requirement can be expressed as 
        \begin{equation}\label{eq:relaxcrit}
            \subsc{\tau}{relax} \la \subsc{\tau}{crit}.
        \end{equation}
        Here, $\subsc{\tau}{crit}$ is the critical thermal relaxation timescale defined by \citep{LinYoudin2015}
        \begin{equation}\label{eq:taucrit}
            \subsc{\tau}{crit} =\frac{H_g}{R}\frac{|q|}{\gamma-1}\Omega_{\rm K}^{-1},
        \end{equation}
        where $\Omega_{\rm K}$ is the Keplerian frequency, $H_g$ is the gas scale height, and $q$ is the radial gradient of the temperature.
        Strictly speaking,  high-$k_x$ unstable modes persist for $\tau_{\rm relax} \ga \tau_{\rm crit}$, but we neglect these modes because their growth rates are much smaller than the maximum growth rates for $\tau_{\rm relax} \la \tau_{\rm crit}$. 
        We describe more details of the thermal relaxation timescale $\subsc{\tau}{relax}$ in Section \ref{subsubsec:method_unstabel_area}.
        }
        
        { In realistic protoplanetary disks, Equation~\eqref{eq:relaxcrit} is fulfilled in regions around the midplane \citep[][see also Section~\ref{sec:results} of this paper]{Malygin+2017,PfeilKlahr2019}. In this study, we refer to such a region as a VSI zone. 
        }
        
    \subsection{Linear Analysis}\label{subsec:VSI_linear_analysis}
        { A number of previous studies already derived dispersion relations with and without gas cooling and vertical stratification {\citep{UrpinBrandenburg1998,Urpin2003,ArltUrpin2004,NelsonGresselUmurhan2013,Barker:2015tt,LinYoudin2015,Lin:2017aa,LatterPapaloizou2018}}. Here, we follow {\citet{NelsonGresselUmurhan2013,Nelson+2016:erratum}} and derive a local dispersion relation.
        }
        
        { We consider a VSI zone with the finite vertical extent and focus on linear modes whose wavelengths are short enough to fit into the zone (see Section~\ref{subsubsec:method_wavenumber} for the allowed range of wavenumbers). 
        We apply a locally isothermal equation of state to the VSI zone and regard the modes within the zone as radially and vertically local. 
        We note that \citet{LinYoudin2015} presents a dispersion relation for vertically global VSI modes including the effect of finite thermal relaxation.
        However, we do not use this here because it is not strictly applicable to disks with vertically varying $\tau_{\rm relax}$.
        }

        { Under the local shearing box approximation, the equation of continuity can be written as
        \begin{equation}\label{eq:renzoku1}
            \frac{\partial \rho_g}{\partial t} + \nabla \cdot \paren{\rho_g { v}} = 0,
        \end{equation}
        where $\rho_g$ is the gas density and ${ v} = (v_x,~v_y,~v_z)$ is the gas velocity with three components for the radial, azimuthal and vertical velocities in the Cartesian shearing box. The equations of motion of gas in the shearing box are
        \begin{equation}\label{eq:motionx1}
            \frac{\partial v_x}{\partial t} + \paren{{ v}\cdot \nabla}v_x = -\frac{1}{\rho_g}\frac{\partial P}{\partial x} + 2\Omega_0v_y + 3\Omega_0^2x,
        \end{equation}
        \begin{equation}\label{eq:motiony1}
            \frac{\partial v_y}{\partial t} + \paren{{ v}\cdot \nabla}v_y = -\frac{1}{\rho_g}\frac{\partial P}{\partial y}-2\Omega_0v_x,
        \end{equation}
        \begin{equation}\label{eq:motionz1}
            \frac{\partial v_z}{\partial t} + \paren{{ v}\cdot \nabla}v_z = -\frac{1}{\rho_g}\frac{\partial P}{\partial z}-g,
        \end{equation}
        where $\Omega_0 = \mathrm{constant}$ is the angular velocity in the shearing box, and $g = \Omega_0^2z$ is the vertical component of  stellar gravity. 
        Following \citet{GS67}, we consider the vicinity of a given height $z=z_0$ and {assume that $g$ and the isothermal sound speed $c_s = \sqrt{P/\rho_g}$}  are approximately constant on this small vertical scale.
        }

        { We assume axisymmetry and consider perturbations of the form $\propto e^{-i \omega t + i k_x x + i (k_z- ig/(2c_s^2))z }$ on the steady background. Here, $\omega$ is the angular frequency and $k_x$ and $k_z$ are the radial and vertical wavenumbers, respectively. The factor $e^{gz/(2c_s^2)}$ accounts for the variation of the background density in the vertical direction  \citep{NelsonGresselUmurhan2013},  The dispersion relation for the perturbations is {(see \citet{NelsonGresselUmurhan2013} and \citet{Nelson+2016:erratum} for a derivation) }
        \begin{eqnarray}\label{eq:bunsan2}
                \omega^4 - \left[c_s^2\paren{k_x^2 + k_z^2}+\kappa_0^2 + {\frac{g^2}{4c_s^2}}\right]\omega^2 & \nonumber \\
                -2\Omega_0c_s^2k_xk_z\frac{\partial v_{y0}}{\partial z} + i\Omega_0gk_x\frac{\partial v_{y0}}{\partial z} + &\kappa_0^2\paren{c_s^2k_z^2 + {\frac{g^2}{4c_s^2}}} = 0,
        \end{eqnarray}
        where $v_{y0}$ and $\kappa_0$ are the rotation velocity deviation and epicyclic frequency, respectively, for the background; the latter one can be written as
        \begin{equation}\label{eq:kappa0}
            \kappa_0^2 = 2\Omega_0 \paren{\frac{\partial v_{y0}}{\partial x} + 2\Omega_0} .
        \end{equation}
        }

        { For an unstable perturbation that satisfies Equation \eqref{eq:bunsan2}, the growth rate is given by 
        \begin{equation}\label{eq:GammaVSI}
            \subsc{\Gamma}{VSI} = \mathrm{Im}\paren{\omega}.
        \end{equation}
        }

\section{Model}\label{sec:model}
    { We study the VSI linear stability of protoplanetary disks taking into account dust growth and settling. We present a gas disk model in Section \ref{subsec:disk}, a model for calculating the disk relaxation time and VSI growth rate in Section \ref{subsec:unstable}, a dust model used to calculate the thermal relaxation time in Section \ref{subsec:dust}, and our parameter choice in Secrion \ref{subsec:method}.
    }

    \subsection{Gas Disk Model}\label{subsec:disk}
        \begin{figure}[t]
            \includegraphics[width=8cm,bb = 0 0 653 1021]{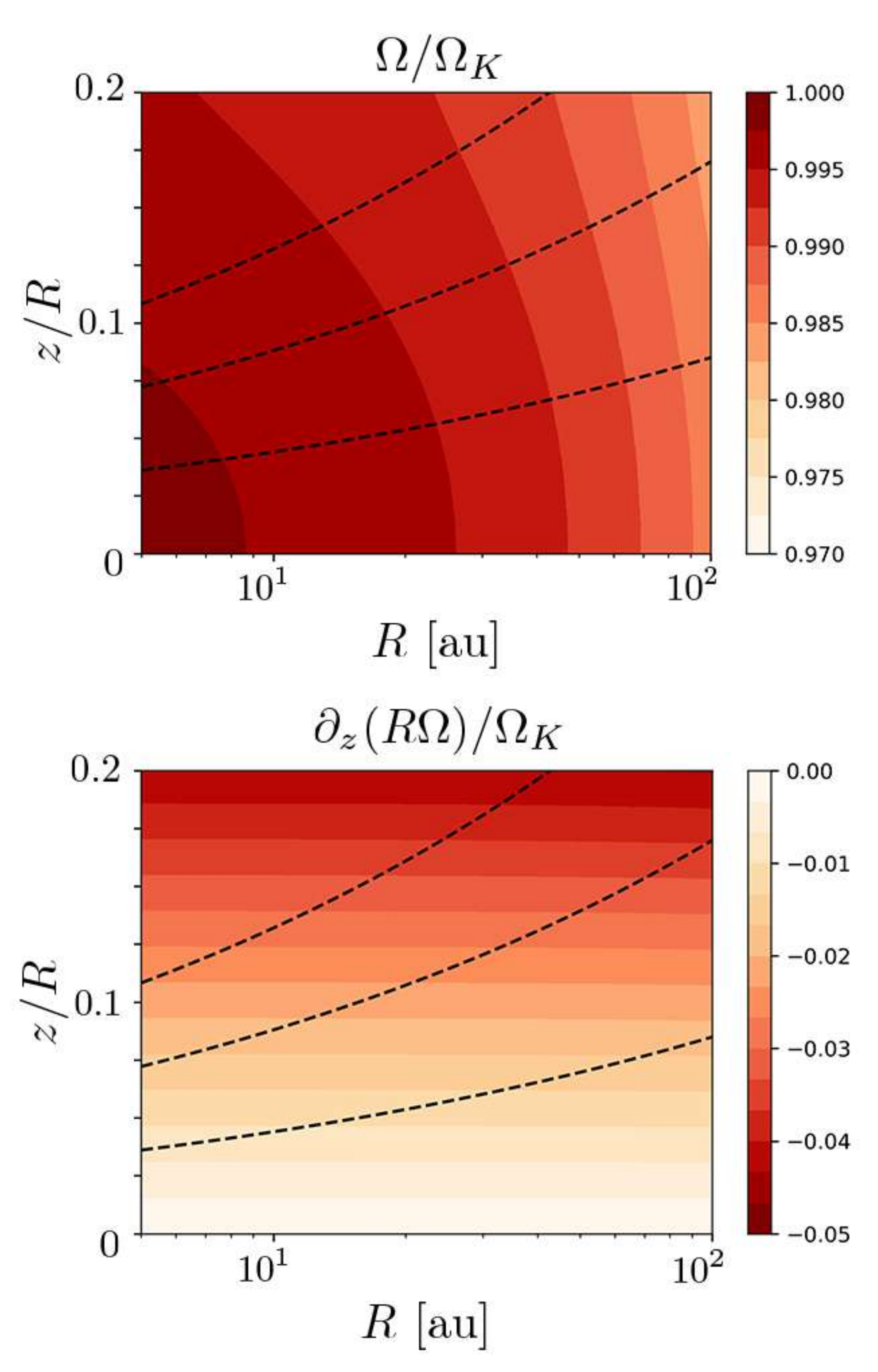}
            \caption{{Local rotational angular velocity $\Omega$ (upper panel) and vertical shear $\pd \paren{R\Omega}/\pd z$ (lower panel), as a function of $R$ and $z/R$. The dashed lines represent $H_g$, $2H_g$, and $3H_g$ in height from the midplane.}
            }
            \label{fig:NzOmegaVertical}
        \end{figure}
    
        { We consider an axisymmetric disk around a solar-mass star.
        The gas surface density is given by
        \begin{equation}\label{eq:Sigmagas}
            \Sigma_g(R) =  \frac{\paren{2-\beta_{\Sigma}}\subsc{M}{disk}}{2\pi R_c^2}\paren{\frac{R}{R_c}}^{-\beta_{\Sigma}} \exp{\left[-\paren{\frac{R}{R_c}}^{2-\beta_{\Sigma}}\right]} ,
        \end{equation}
        where $\subsc{M}{disk}$ is the total mass of the gas disk, $R_c$ is the characteristic radius, and $\beta_\Sigma$ is a dimensionless number characterizing the radial slope of the gas surface density. 
        Equation~\eqref{eq:Sigmagas} is motivated by the similarity solution of the viscous accretion disk model \citep{Lynden-Bell:1974aa,Hartmann:1998aa} although we do not consider disk evolution in this study. 
        Since we are primarily interested in the effects of dust growth and settling on the VSI, we fix the gas disk parameters to $\subsc{M}{disk} = 0.01M_\sun$, $R_c = 100~ \au$, and $\beta_\Sigma = 1$ except in  Section~\ref{subsec:result_mass}, where we show that disks of higher disk dust masses have more extended VSI zones.  
        The heat capacity ratio $\gamma$ and mean molecular mass $m_g$ of the gas is taken to be 1.4 and $2.3m_p$, respectively, where $m_p$ is the proton mass.
        }

        { We focus on the outer region of the disk where the temperature is determined by stellar irradiation.
        Assuming that the disk is optically thick to stellar radiation \footnote{ A disk region that is optically thin to its own thermal emission can be optically thick to the radiation from the central star because the opacity in the visible is higher than that in the infrared and, more importantly, because the radial optical depth is $\sim R/H_g$ ($\sim 10$--100) times larger than the vertical optical depth \citep{Chiang:1997aa}.} and that the stellar luminosity is equal to the solar luminosity, the temperature of the disk interior is given by
        \begin{equation}\label{eq:Tgas}
            T(R) = 130\paren{\frac{R}{1{\rm ~au}}}^q {\rm ~K}
        \end{equation}
        with $q = -3/7$  \citep{Chiang:1997aa}. Viscous heating is negligible as long as we focus on $R \ga 10~\rm au$ \citep[e.g.,][]{Bitsch:2015aa}. We assume that the disk interior is vertically isothermal, neglecting warmer surface layers that are optically thin to the starlight \citep{Chiang:1997aa}. 
        }

        { From vertical hydrostatic equilibrium, the gas density is given by
        \begin{equation}\label{eq:rhogas}
            \rho_g(R,~z) = \frac{\Sigma_g}{\sqrt{2\pi}H_g} \exp{\left(-\frac{z^2}{2H_g^2}\right)} 
        \end{equation}
        with $H_g = c_s/\Omega_{\rm K}$, $\Omega_{\rm K} = \sqrt{GM_{\ast}/R^3}$, and $M_* = 1M_\sun$. 
        }

        { The vertical shear of the gas rotation velocity, $\partial v_{y0} /\partial z= \partial(R\Omega)/\partial z $,  characterizes the strength of the VSI.
        Because we assume a radially varying temperature profile, $\partial(R\Omega)/\partial z$ 
        is nonzero at all height except at the midplane. 
        This can be analytically shown for $R \ll R_c$, where $\Sigma_g \propto R^{-\beta_\Sigma}$. In this region, $\Omega$ and $\partial(R\Omega)/\partial z$ can be approximated as \citep{TakeuchiLin2002}
        \begin{equation}\label{eq:Omega2}
           \Omega\paren{R,~z} \approx \Omega_{\rm K} \left[ 1+\frac{1}{2}\paren{\frac{H_g}{R}}^2 \paren{-\beta_{\Sigma}{  +\frac{q-3}{2}}+\frac{q}{2}\frac{z^2}{H_g^2}} \right],
        \end{equation}
        and 
        \begin{equation}\label{eq:vertical_shear2}
           \frac{\pd \paren{R\Omega}}{\pd z} \approx \frac{q}{2}\frac{z}{R}\Omega_{\rm K},
        \end{equation}    
        respectively.
        Equation~\eqref{eq:vertical_shear2} proves that a nonzero radial temperature gradient is the source of the vertical shear \citep[e.g.,][]{LinYoudin2015}. 
        Figure \ref{fig:NzOmegaVertical} shows the gas angular velocity $\Omega(R,z)$ and the vertical shear of the angular velocity $\partial(R\Omega)/\partial z$ in the disk, including the region $R \ga R_c$. From the {lower} panel of Figure \ref{fig:NzOmegaVertical}, and also from Equation~\eqref{eq:vertical_shear2}, the vertical shear increases with $z$, suggesting that the VSI is stronger at higher altitude as long as the cooling criterion (Equation~\eqref{eq:relaxcrit}) is fulfilled.
        }
        
        {
        We assume that the disk is weakly turbulent and express the turbulent diffusion coefficient as $\alpha_Dc_sH_g$, where $\alpha_D$ is the dimensionless parameter characterizing the level of turbulent diffusion. 
        Turbulent diffusion controls the maximum wavenumber of the VSI modes (Section~\ref{subsubsec:method_wavenumber}) and the vertical scale height of dust particles (Setion~\ref{subsec:dust}). 
        In principle, $\alpha_D$ in outer disk regions should depend on the strength of VSI-driven turbulence, and hence on the VSI growth rate $\subsc{\Gamma}{VSI}$, which is the output of our model. 
        Therefore, a self-consistent determination of $\alpha_D$ and $\subsc{\Gamma}{VSI}$ requires a model that predicts the former as a function of the latter. 
        Lacking such a model, we opt for taking $\alpha_D$ as a free parameter, although we do discuss potential feedback of dust settling on the level of VSI-driven turbulence in Section \ref{subsec:discuss_feedback}.
        }
       
    \subsection{Instability Analysis}\label{subsec:unstable}
        { We analyze the linear stability of the model disk against the VSI in the following three steps. 
        In the first step, we use the thermal relaxation criterion (Equation~\eqref{eq:relaxcrit}) to search for the disk region, which we call the VSI zone, where the linear VSI operates (Section~ \ref{subsubsec:method_unstabel_area}). 
        In the second step, we compute the range of wavenumbers for the VSI modes that fit into the active zone (Section~\ref{subsubsec:method_wavenumber}). 
        In the third step, we calculate the maximum growth rate of the VSI at each radial location (Section~\ref{subsubsec:method_maximum_growth_rate}).
        }

        \subsubsection{Defining the VSI Zone}\label{subsubsec:method_unstabel_area}

            { We compute the radial and vertical extent of the VSI zone by applying the thermal relaxation criterion (Equation~\eqref{eq:relaxcrit}) to each point ($R$, $z$) in the disk. 
            In protoplanetary disks, local thermal relaxation is regulated either by collisional heat transfer from gas to dust or by radiative cooling \citep{Malygin+2017}. The area around the boundary of the VSI zone can be regarded as optically thin \citep{Malygin+2017}. Following \citet{Malygin+2017} and \citet{PfeilKlahr2019}, we estimate the local thermal relaxation timescale
            $\subsc{\tau}{relax}$ as
            \begin{equation}\label{eq:taurelax}
                \subsc{\tau}{relax} = \max{\paren{\subsc{\tau}{coll},\subsc{\tau}{emit}}},
            \end{equation} 
            where $\subsc{\tau}{coll}$ and $\subsc{\tau}{emit}$ are the timescales of collisional heat transfer and radiative cooling, respectively.
            }

            { The timescale of collisional heat transfer is given by
            \begin{equation}\label{eq:taucoll}
                \subsc{\tau}{coll} = \frac{\subsc{\ell}{gd}}{ \subsc{v}{th}},
            \end{equation}
            where $\subsc{\ell}{gd}$ is the mean travel length of gas molecules before colliding with dust particles and $\subsc{v}{th}$ is the mean relative velocity between the gas molecules and dust particles. The relative velocity $\subsc{v}{th}$ can be approximated as the mean thermal speed of the molecules, 
            \begin{equation}\label{eq:vth}
                \subsc{v}{th} \approx \sqrt{\frac{8\kB T}{\pi m_g}},  
            \end{equation}
            where $\kB$ is the Boltzmann constant. The radiative cooling timescale $\subsc{\tau}{emit}$ in the optically thin limit is given by \citep{Malygin+2017}
            \begin{equation}\label{eq:tauemit}
                \subsc{\tau}{emit} = \frac{C_V}{16\kappa_{\rm P}(T)\subsc{\sigma}{SB} T^3},
            \end{equation}
            where $\kappa_{\rm P}(T)$ is the Planck mean opacity per unit gas mass and $\subsc{\sigma}{SB}$ is the Stefan--Boltzmann constant. Both $\subsc{\ell}{gd}$ and $\kappa_{\rm P}$ depend on the local size distribution of the dust particles. Our dust model is described in Section~\ref{subsec:dust}.
            }
            
            { In general, collisional heat transfer regulates the cooling timescale at high altitude where the dust density is low \citep{Malygin+2017}. 
            Because the dust density decreases monotonically with $|z|$, there is the height $|z| = L$ above which the VSI is stable. 
            In other words, the VSI zone refers to the region where $|z| < L$. Note that $L$ generally depends on $R$ and becomes zero where the VSI is stable at all heights.
            }

        \subsubsection{Wavenumber Range of the VSI Modes}\label{subsubsec:method_wavenumber}
            { The VSI modes generally have $k_xk_z <0$ \citep{ArltUrpin2004}, so we restrict $k_x > 0$ and $k_z < 0$. Below we further restrict the wavelength range permitted  for VSI modes by accounting for viscous damping and the finite thickness of the VSI zone. 
            }

            { Viscous damping erases unstable modes of short wavelengths, giving upper limits on $k_x$ and $|k_z|$ for the true VSI modes. We particularly focus on the upper limit on the radial wavenumbers because VSI modes typical have $k_x \gg |k_z|$ \citep{ArltUrpin2004}.
            The maximum wavenumber set by viscous damping can be estimated as \citep{LinYoudin2015}
            \begin{equation}
                (k_{x,{\rm max}}H_g)^2 \approx \frac{|q|}{\alpha_D}\frac{H_g}{R}
            \end{equation}
            or equivalently, 
            \begin{equation}\label{eq:kxmax}
                k_{x,{\rm max}} \approx \sqrt{\frac{|q|}{\alpha_D R H_g}}.
            \end{equation}
            
            Because the VSI zone has a finite vertical extent, only modes whose vertical wavelengths are short enough to ``fit'' into the zone can be unstable.     
            Specifically, we require the VSI modes at each $R$ to have half wavelengths shorter than the vertical zone width $2L$. In other words, the vertical wavenumbers of the VSI modes must satisfy $|k_z| > |k_z|_{\rm min}$, where
            \begin{equation}\label{eq:kzmin}
                |k_{z}|_{\rm min} \approx \frac{\pi}{2L}.
            \end{equation}
            }

        \subsubsection{The Maximum Growth Rate}\label{subsubsec:method_maximum_growth_rate}
            { The local dispersion relation (Equation \eqref{eq:bunsan2}) formally gives the growth rate of the VSI at each ($R$, $z$). 
            However, the most unstable VSI modes typically have vertical wavelengths comparable to $H_g$ \citep[e.g.,][see also Section \ref{subsec:result_maximum_growth}]{NelsonGresselUmurhan2013}. 
            Physically, such long-wavelength modes should be regarded as extending over the whole vertical extent of the VSI zone.
            }

            { For this reason, we opt for a semi-local approach in which we calculate the maximum VSI growth rate $\Gamma_{\rm VSI,maax}$ at each $R$ using the linear dispersion relation, but limiting the range of vertical wavenumbers to $|k_z| \geqslant |k_z|_{\rm min}$ to account for the finite vertical extent of the VSI zone. 
            The vertically local quantities $g(=-\Omega_{\rm K}^2 z)$ and $ v_{y0}(=R\Omega(R,z))$ involved in the local dispersion relation are evaluated at the vertical boundary of the VSI zone, $|z| = L$, because the VSI is generally more vigorous at higher altitude unless buoyancy suppresses it \citep{NelsonGresselUmurhan2013}.
            We also limit radial wavenumbers to $k_x \leqslant k_{x,{\rm max}}$ to account for viscous damping.
            }
            
            {
            Besides, the local analysis in this study and the vertically global analysis by \citet{LinYoudin2015} give similar predictions for the body mode in a vertically wide VSI zone (see Section~\ref{subsec:discuss_appropriate}).
            }

    \subsection{Dust Model}\label{subsec:dust}
        { We here describe the dust model we use to calculate $\ell_{\rm dg}$ and $\kappa_{\rm P}$. 
        We consider spherical, icy dust particles of internal density $\subsc{\rho}{int} = 1 {\rm ~g~cm^{-3}}$. 
        The ratio between the dust surface density $\Sigma_d$ and $\Sigma_g$ is fixed to the interstellar dust abundance of $1\%$, whereas the local dust-to-gas ratio is allowed to vary with $z$ considering dust settling. As we discuss in Section~\ref{subsec:result_mass}, the VSI zone shrinks as  $\Sigma_d$ decreases.
        }
        
        { The particle size distribution is assumed to follow a power law
        \begin{eqnarray}\label{eq:Sigmad_a}
            \frac{dN_{d}(a)}{da} =
            \left\{
            \begin{array}{ll}
            \frac{3\Sigma_d}{8\pi \rho_{\rm int}\paren{\sqrt{\subsc{a}{max}}-\sqrt{\subsc{a}{min}}}}a^{-7/2}, & \subsc{a}{min} < a < \subsc{a}{max}, \\
                0, & {\rm otherwise}, 
            \end{array}
            \right. 
        \end{eqnarray}
        where $dN_{d}(a)/da$ is the number surface density per unit particle size $a$, $\Sigma_d$ is the total dust mass surface density, and $\subsc{a}{min}$ and $\subsc{a}{max}$ are the minimum and maximum particle sizes, respectively.
        The size distribution given by Equation~\eqref{eq:Sigmad_a} satisfies the normalization
        \begin{equation}\label{eq:Sigmad}
            \Sigma_d = \int_{\subsc{a}{min}}^{\subsc{a}{max}} m\frac{dN_{d}(a)}{da} da,
        \end{equation}
        where $m = (4\pi/3)\rho_{\rm int}a^3$ is the particle mass.
        The power-law slope assumed in Equation~\eqref{eq:Sigmad_a} is simply taken from the interstellar particle size distribution \citep{Mathis:1977aa}. 
        We note, however, that the size distribution can be somewhat shallower or steeper than assumed in Equation \eqref{eq:Sigmad_a} depending on the details of collisional growth and fragmentation \citep{Birnstiel:2011aa}. 
        }
        
        { We take the maximum particle size $\subsc{a}{max}$ as a free parameter to study the impact of dust growth on the VSI. 
        The minimum particle size is less well defined, but we may crudely taken it to be $\sim$ 0.1--1 $\micron$ because particles smaller than these sizes grow quickly through Brownian motion \citep{Birnstiel:2011aa}. We fix $\subsc{a}{min} = 1~\micron$ throughout this study. 
        }

        { Assuming the balance between settling and diffusion, the vertical distribution of the particles can be written as 
        \citep{TakeuchiLin2002}
        \begin{equation}\label{eq:rhoda}
            \frac{dn_d(a,z)}{da} = C_d(a) \exp{\left[-\frac{z^2}{2H_g^2}-\frac{{\rm St}_{{\rm mid}}}{\alpha_D}\paren{\exp\frac{z^2}{2H_g^2}-1} \right]},
        \end{equation}
        where $dn_d(a,z)/da$ is the particle number density per unit radius at height $z$, $\mathrm{St}_{{\rm mid}}$ is the Stokes number of the particles at the midplane, and $C_d(a)$ is the normalized constant determined by the condition
        \begin{equation}\label{eq:C_d}
           \frac{dN_d(a)}{da} = \int \frac{dn_d(a,z)}{da} dz.
        \end{equation} 
        The Stokes number is the product of the stopping time and Keplerian frequency. Assuming that the particle radius are smaller than the mean free path of the disk gas molecules, gas drag onto the particles follows Epstein's law, which gives \citep[see, e.g., ][]{Birnstiel+2010}
        \begin{equation}\label{eq:St}
            \mathrm{St}_{{\rm mid}} = \frac{\pi \subsc{\rho}{int} a}{2\Sigma_g}.
        \end{equation}
        To evaluate $C_d$, we note that the integral in Equation \eqref{eq:C_d} is dominated by the region $z \la H_g$, for which the Equation~\eqref{eq:rhoda} can be approximated as ${dn_d(a,z)}/{da} \approx C_d(a)\exp(-z^2/(2H_d^2))$, where 
        \begin{equation}\label{eq:Hd}
            H_{d} = \paren{1+\frac{\mathrm{St}_{{\rm mid}}}{\alpha_D} }^{-1/2}H_g
        \end{equation}
        represents the scale height of particles with size $a$ \citep{Dubrulle+1995,YoudinLithwick2007}. This approximation gives
        \begin{equation}\label{eq:C_d2}
           C_d(a) = \frac{1}{\sqrt{2\pi}H_d}\frac{dN_{d}(a)}{da}.
        \end{equation}
        }

        { The vertical--size distribution $dn_d(a,z)/da$ gives $\subsc{\ell}{gd}$ and $\kappa_{\rm P}$ as a function of $z$. The mean travel length of gas molecules colliding with dust particles $\subsc{\ell}{gd}$ is given by 
        \begin{equation}\label{eq:l_dg}
            \subsc{\ell}{gd} = \left(\int_{\subsc{a}{min}}^{\subsc{a}{max}}\pi a^2\frac{dn_d}{da}  da\right)^{-1}.
        \end{equation}
        The Planck mean opacity $\kappa_{\rm P}$ per unit gas mass can be written as
        \begin{equation}\label{eq:opacity}
            \kappa_{\rm P}(T)= \frac{1}{\rho_g}\int_{\subsc{a}{min}}^{\subsc{a}{max}}\sigma_{\rm abs,P}(a,T)\frac{dn_d}{da} da,
        \end{equation}
        where $\sigma_{\rm abs,P}(a,T)$ is the Planck mean absorption cross section of the particles.
        In this study, we approximate $\sigma_{\rm abs,P}$ with the monochromatic absorption cross section $\sigma_{\rm abs}(a,\lambda)$ at wavelength $\lambda = \lambda_{\rm peak}(T)$, where
        \begin{equation}\label{eq:lambda}
            \subsc{\lambda}{peak}(T) = 10{\rm ~\mu m} \paren{\frac{300{\rm ~K}}{T}}
        \end{equation}
        is the peak wavelength of the Planck function.
        Furthermore, we crudely approximate $\sigma_{\rm abs}(a,\lambda)$ as
        \begin{equation}\label{eq:sigmaabs}
            \subsc{\sigma}{abs}(a,\lambda) = \pi a^2 \min\left(1,\frac{2\pi a}{\lambda}\right),
        \end{equation}
        where the factor $2\pi a/\lambda$ is called the size parameter in Mie scattering theory \citep[see, e.g.,][]{Bohren:1983aa}. 
        This approximate expression satisfies the asymptotic properties of Mie scattering that $\subsc{\sigma}{abs}$ approaches the geometric cross section $\pi a^2$ in the short wavelength limit ($\lambda \ll 2\pi a$) and that the dust mass opacity $\subsc{\sigma}{abs}/m$ is independent of $a$ in the long wavelength limit ($\lambda \gg 2\pi a$).
        }
        
        {
        We neglect the radial inward drift of the dust particles due to gas drag \citep{Whipple:1972vv,Adachi:1976uv,Weidenschilling:1977wt} and assume that the dust surface density profile is constant in time. 
        This approach is valid if the VSI growth timescale is shorter than the dust drift timescale. 
        For ${\rm St}_{\rm mid} < 1$, which is the case for the dust particles considered in our model, the drift timescale can be estimated as $\sim (R/H_g)^2 {\rm St}_{\rm mid}^{-1} \Omega_{\rm K}^{-1}$ (see the references listed above).  
        In the limit of short relaxation timescales,
        the VSI grows on a timescale of $\sim (R/H_g)\Omega_{\rm K}^{-1}$ \citep[e.g.,][]{UrpinBrandenburg1998,NelsonGresselUmurhan2013},
        which is shorter than the drift timescale as long as ${\rm St}_{\rm mid} < 1$. 
        However, radial dust drift may not be negligible when a finite relaxation time suppresses the growth of the VSI.
        In this case, dust depletion due to the radial inward drift would further suppress the VSI as we demonstrate in Section~\ref{subsec:result_mass}. 
        }
        
    \subsection{Computational Domain and Parameter Choices}\label{subsec:method}
        { We consider a disk region defined by $5{\rm ~au} < R < 100{\rm ~au}$ and $ 0 < z/R < 0.2$ and divide the region into 1000 logarithmically spaced radial grids and 1000 linearly spaced vertical grids. 
        The dust size distribution is divided into logarithmic bins of 10 grids per decade in $a$. 
        The main free parameters in our model are the vertical diffusion coefficient $\alpha_D$ and maximum dust particle size $\subsc{a}{max}$. 
        We take $\alpha_D = 10^{-3},10^{-4}$, and $10^{-5}$ and $\subsc{a}{max} = 10 ~\micron,100~\micron$, and $1{\rm ~mm}$.
        }
    
\section{Results}\label{sec:results}
    \begin{figure}[t]
        \includegraphics[width=8cm,bb = 0 0 653 1021]{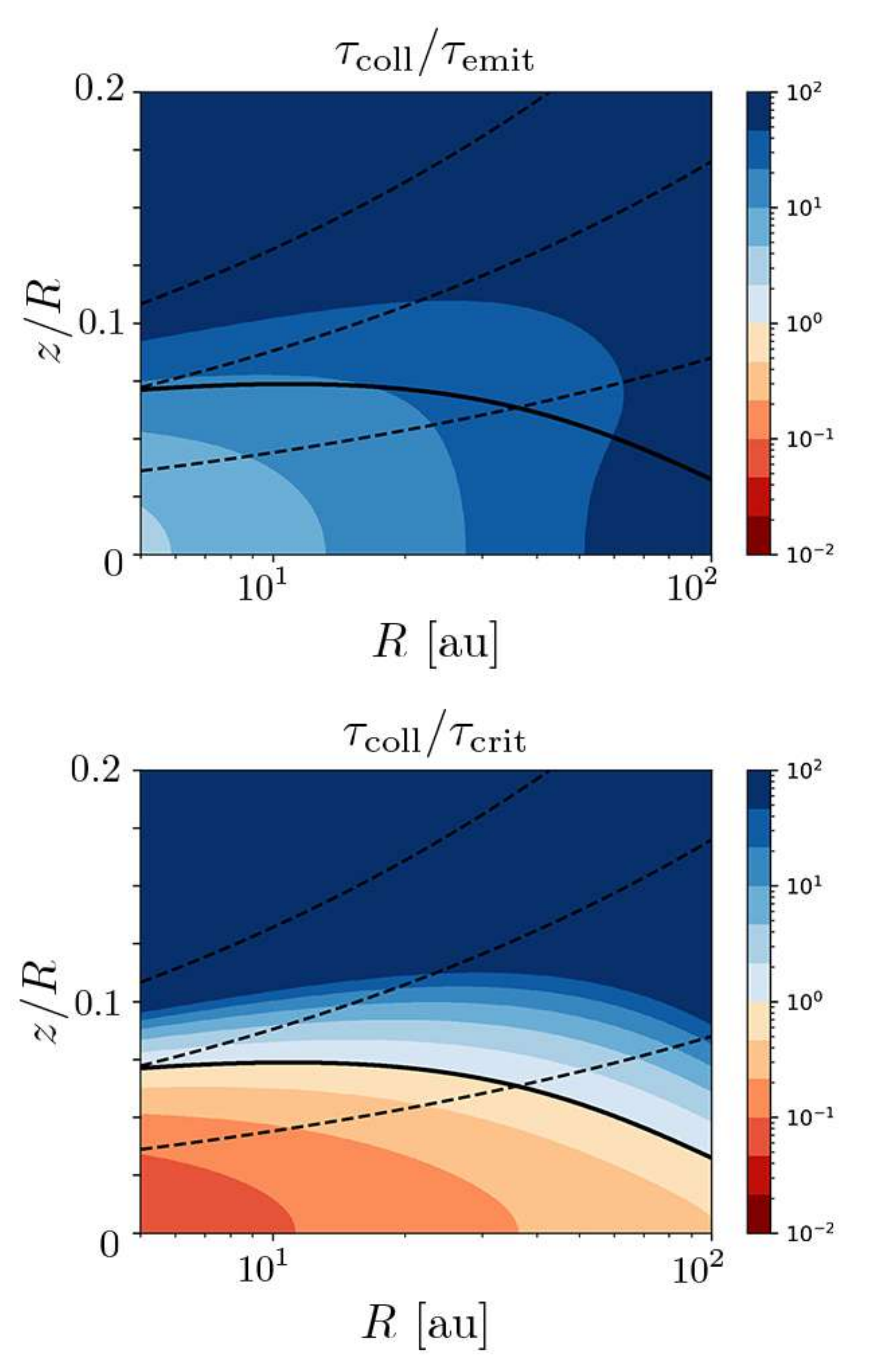}
        \caption{Collisional timescale $\subsc{\tau}{coll}$ normalized by radiative cooling timescale $\subsc{\tau}{emit}$ (upper panel) and critical thermal relaxation timescale $\subsc{\tau}{crit}$ (lower panel) as a function of $R$ and $z/R$ for $\alpha_D=10^{-4}$ and $\subsc{a}{max} = 10~\micron$. The thermal relaxation timescale is given by the larger of the two (see Equation~\eqref{eq:taurelax}). The dashed lines represent $z = H_g$, $2H_g$, and $3H_g$. The solid line marks $\subsc{\tau}{coll}=\subsc{\tau}{crit}$.
        }
        \label{fig:emit_coll}
    \end{figure}
    { In this section, we use the model presented in Section~\ref{sec:model} to study how dust growth and settling affects the VSI stability of protoplanetary disks. 
    We map the VSI zones in Section~\ref{subsec:result_unstable_area} and then compute the radial distribution of the growth rate and wavenumbers of the most unstable VSI model in Section \ref{subsec:result_maximum_growth}.
    We study the dependence of the VSI zone size on the dust and gas surface densities in Section~\ref{subsec:result_mass}.
    }
    
    \subsection{Radial and Vertical Extent of the VSI Zone }\label{subsec:result_unstable_area}
    
        \begin{figure*}
            \centering
            \includegraphics[width = 18cm,bb = 0 0 1815 512]{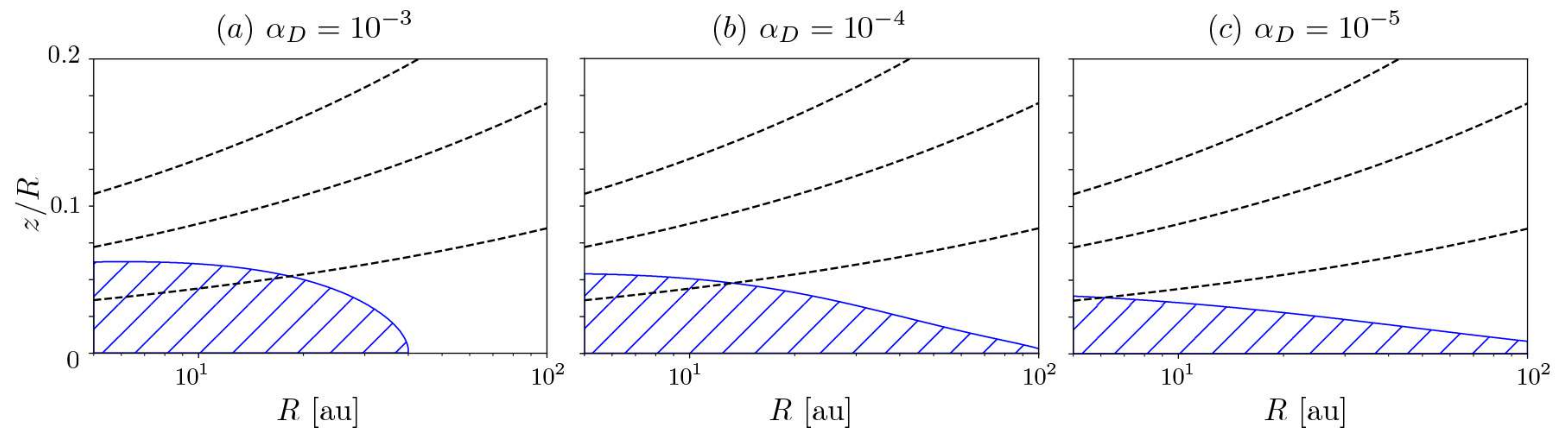}
            \caption{Location of the VSI zone (the shaded area)  for different values of $\alpha_D$ with $\subsc{a}{max} = 100~\micron$. The dashed lines represent $H_g$, $2H_g$, and $3H_g$ in height from the midplane.}
            \label{fig:VSIarea_amax1e-2_alpha}
        \end{figure*}
        \begin{figure*}
            \centering
            \includegraphics[width = 18cm,bb = 0 0 1525 512]{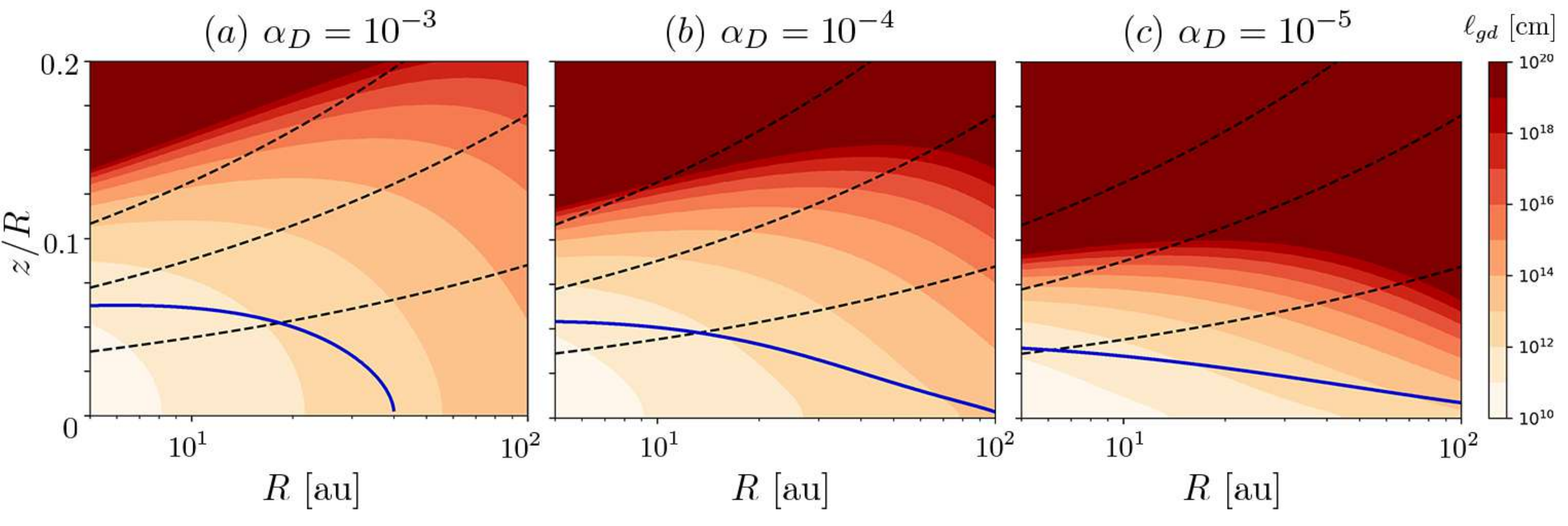}
            \caption{Mean travel length of gas molecules before colliding with dust particles, $\subsc{\ell}{gd}$, as a function of $R$ and $z/R$ for different values of $\alpha_D$ with $\subsc{a}{max}=100~\micron$. The dashed lines represent $H_g$, $2H_g$, and $3H_g$ in height from the midplane. The solid lines correspond to the boundary of the VSI zone.}
            \label{fig:Atot_amax1e-2_alpha}
        \end{figure*}
        
        { 
        As described in Section~\ref{sec:VSI}, the VSI is active where the thermal relaxation time $\subsc{\tau}{relax}$ (Equation~\eqref{eq:taurelax}) is shorter than the critical timescale $\tau_{\rm crit}$ (Equation~\eqref{eq:taucrit}). 
        We find that  $\subsc{\tau}{relax}$ (Equation~\eqref{eq:taurelax}) is determined by the collisional cooling timescale $\subsc{\tau}{coll}$ for all parameters and all regions explored in this study.
        Figure \ref{fig:emit_coll} plots $\tau_{\rm coll}/\tau_{\rm coll}$ and $\tau_{\rm coll}/\tau_{\rm crit}$ as a function of $R$ and $z$ for $\alpha_D=10^{-4}$ and $\subsc{a}{max} = 10~\micron$, showing that $\subsc{\tau}{coll}$ is about an order-of-magnitude larger than $\subsc{\tau}{emit}$ at all locations. 
        Our result is consistent with that of \cite{Malygin+2017}, who showed that $\subsc{\tau}{coll} > \subsc{\tau}{emit}$ in an optically thin region away from the central star (see their Figure~3). 
        The maps of $\tau_{\rm relax}$ (in units of $\Omega_{\rm K}^{-1}$) for all parameter sets are shown in Figures \ref{fig:coll_alpha} and \ref{fig:coll_amax} in Appendix \ref{appendix:timescale}. 
        
        The solid line in the lower panel of Figure~\ref{fig:emit_coll} indicates the boundary of the the VSI zone; below this line, one has $\subsc{\tau}{relax} (=\subsc{\tau}{coll}) < \subsc{\tau}{crit}$ and the VSI operates.\footnote{  
        To be precise, the VSI zone does not include the midplane ($z=0$), where the vertical shear vanishes.
        }
        In this example, the VSI zone extends to $z \sim $0.5--$2H_{\rm g}$ at $5~\au<R<30~\au$. 
        The vertical width of the VSI zone diminishes as $R$ increases.
        The vertical optical depth from infinity to the VSI zone boundary is $O(1)$ at $R \sim 10~\rm au$ and decreases with increasing $R$, so using the thermal relaxation timescale for the optically thin limit (Section \ref{subsubsec:method_unstabel_area}) is marginally justified.
        }
        
        \begin{figure*}[t]
            \centering
            \includegraphics[width = 18cm,bb = 0 0 1815 512]{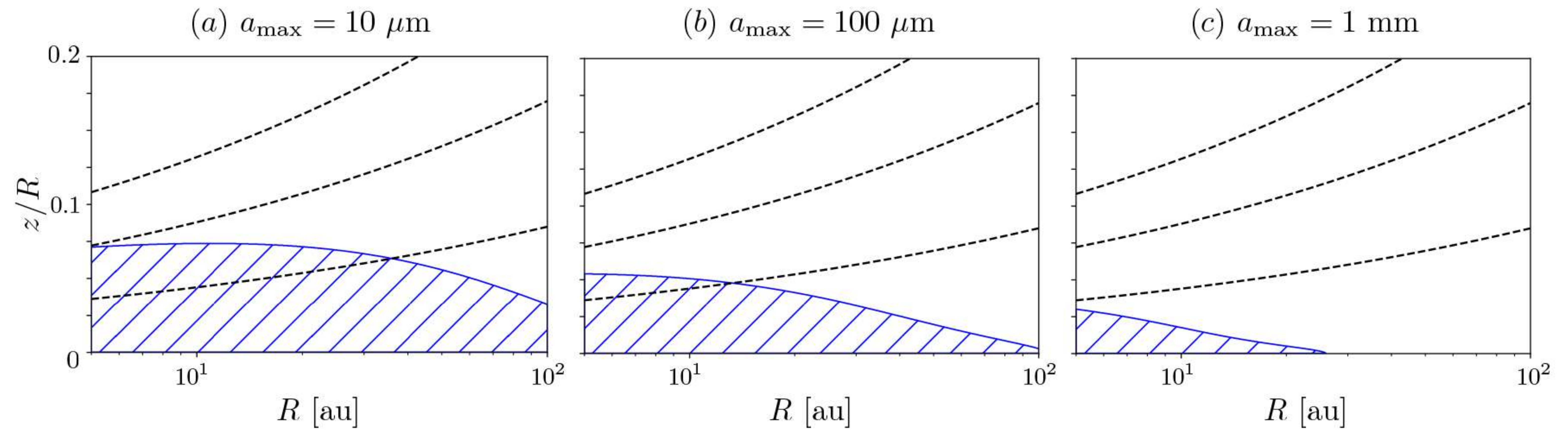}
            \caption{Location of the VSI zone (the shaded area) for different values of $\subsc{a}{max}$ with $\alpha_D = 10^{-4}$. The dashed lines represent $H_g$, $2H_g$, and $3H_g$ in height from the midplane.}
            \label{fig:VSIarea_alpha1e-4_amax}
        \end{figure*}
        \begin{figure*}
            \centering
            \includegraphics[width = 18cm,bb = 0 0 1525 512]{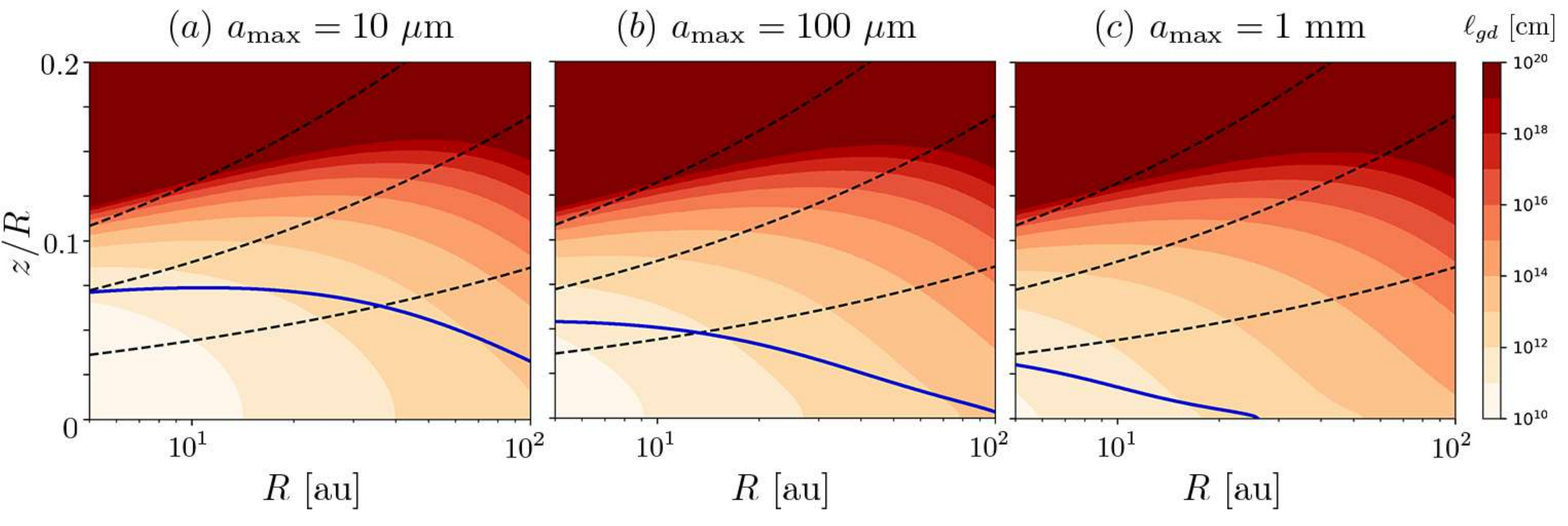}
            \caption{Mean travel length of gas molecules before colliding with dust particles, $\subsc{\ell}{gd}$, as a function of $R$ and $z/R$ for different values of $\subsc{a}{max}$ with $\alpha_D = 10^{-4}$. The dashed lines represent $H_g$, $2H_g$, and $3H_g$ in height from the midplane. The solid lines correspond to the boundary of the VSI zone. }
            \label{fig:Atot_alpha1e-4_amax}
        \end{figure*}
        
        { Figure \ref{fig:VSIarea_amax1e-2_alpha} indicates the location of the VSI zone for $a_{\rm max} = 100~\rm \mu m$ but with different values of $\alpha_D$. 
        This figure illustrates how dust settling affects the extent of the VSI zone; in our model, the dust scale height decreases with decreasing $\alpha_D$ (see Equation~\eqref{eq:Hd}). 
        We find that dust settling leads to a VSI zone that is more confined to the midplane region and more extended to larger radial distances.
        This is because the settling causes dust depletion and dust concentration, which increase and decrease the thermal relaxation timescale $\tau_{\rm relax} = \tau_{\rm coll} \propto \ell_{\rm gd}$, at high and low altitude, respectively { (see also Figure \ref{fig:coll_alpha})}.
        See Figure \ref{fig:Atot_amax1e-2_alpha} for the variation of $\ell_{\rm gd}$ at different locations with $\alpha_D$.
        }
        
        {The stabilizing effect of dust setting on the VSI at the midplane can also be confirmed by looking at how $\ell_{\rm gd}$ at the midplane depends on $\alpha_{\rm D}$.
        For simplicity, we assume $\alpha_D < {\rm St}_{{\rm mid}}$, which holds in our model at sufficiently large $R$ ($R \gtrsim 50 ~\au$ for $\alpha_D=10^{-4}$). 
        With this assumption, we approximate $H_d \sim (\alpha_D/\subsc{\mathrm{St}}{mid})^{1/2} H_g$ and $dn_d(z=0)/da \propto ({\rm St}_{\rm mid}/\alpha_D)^{1/2}dN_d/da$.
        Using this, the integration in Equation \eqref{eq:l_dg} can be analytically performed, resulting in 
        \begin{equation}\label{eq:ldg_approx}
           \subsc{\ell}{gd}(z=0) \propto \frac{\Sigma_{\rm g}^{1/2}}{\Sigma_{\rm d}} \frac{\subsc{a}{max}^{1/2}}{\log\paren{\subsc{a}{max}}} \alpha_D^{1/2}
        \end{equation}
        for $\subsc{a}{max} \gg \subsc{a}{min}$.
        Equation \eqref{eq:ldg_approx} confirms that  $\subsc{\ell}{gd}$ at the midplane decreases with decreasing $\alpha_D$ (see Figure~\ref{fig:Atot_amax1e-2_alpha}). 
        }

        { Figure \ref{fig:VSIarea_alpha1e-4_amax} indicates the location of the VSI zone for different values of $a_{\rm max}$, illustrating how the VSI zone evolves with dust growth. The figure shows that the VSI zone shrinks toward the midplane and toward the central star as $\subsc{a}{max}$ increases.
        This is because increasing $a_{\rm max}$ decreases dust particles' total surface area and thus increases $\ell_{\rm gd}$ as shown in Figure \ref{fig:Atot_alpha1e-4_amax}.
        In fact, increasing $a_{\rm max}$ also promotes dust settling, which acts to decrease $\ell_{\rm gd}$ at the midplane.
        However, we find that this effect is minor compared to the increase in the midplane $\ell_{\rm gd}$ due to local dust growth. This can also be confirmed from Equation \eqref{eq:ldg_approx} implying that $\ell_{\rm gd}$ at the midplane increases with $\subsc{a}{max}$.
        Well above the midplane, both dust settling and local dust growth increase $\subsc{\ell}{gd}$, stabilizing the VSI. 
        A closer inspection shows that local dust growth dominates the increase of $\ell_{\rm gd}$ at $z \la 1$--$2H_g$. 
        }
        
    \subsection{The Maximum Growth Rate and Wavenumbers}\label{subsec:result_maximum_growth}
        \begin{figure}[t]
            \includegraphics[width = 8cm,bb = 0 0 781 570]{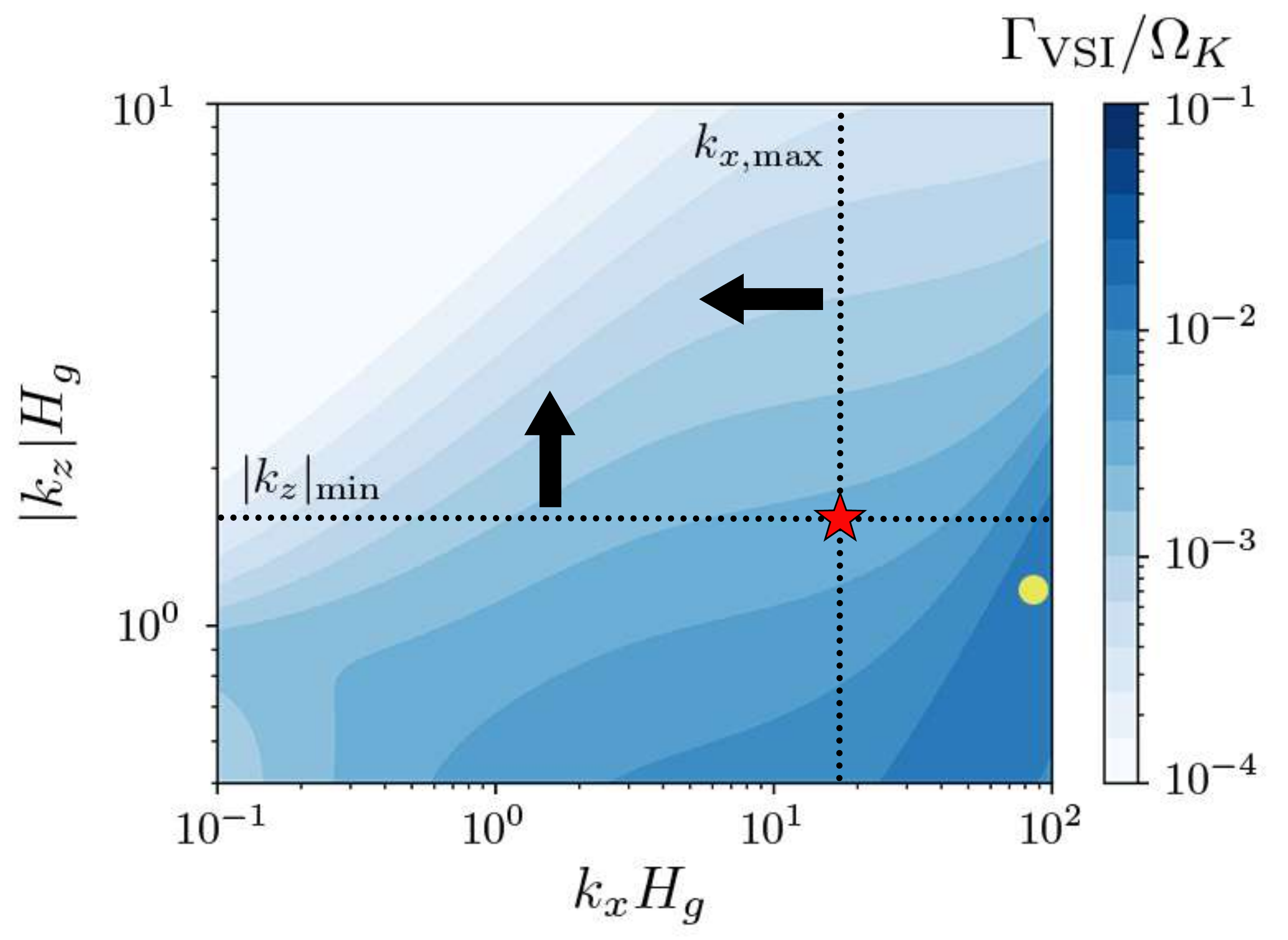}
            \caption{
            VSI growth rate  $\subsc{\Gamma}{VSI}$  as a function of $k_x$ and $|k_z|$ at the top  of the VSI zone at $R = 60~\au$ for $\subsc{a}{max}=10~\micron$ and $\alpha_D = 10^{-4}$. The vertical and horizontal dotted lines indicate $k_x = k_{x,{\rm max}}$ and $k_z = |k_z|_{\rm min}$, respectively. The star symbol marks the most unstable VSI mode in the range $k_x<k_{x,\max}$ and $|k_z| > |k_z|_{\rm min}$, which is limited by viscous damping and the finite vertical extent of the VSI zone. For comparison, the filled circle marks the mode that would be most unstable if the ranges of $k_x$ and $k_z$ were not limited. 
            }
            \label{fig:VSI_colormap_amax1e-3_alpha1e-4}
        \end{figure}

        {
        We search for the most unstable VSI mode at each $R$ using the procedure described in Section~\ref{subsubsec:method_maximum_growth_rate}. 
        For the entire parameter space explored in this study, we find that the must VSI mode always lies at $k_x = k_{x,\rm max}$ and $|k_z| = |k_{z}|_{\rm min}$. 
        This is illustrated in Figure~\ref{fig:VSI_colormap_amax1e-3_alpha1e-4}, where we plot the VSI growth rate $\subsc{\Gamma}{VSI}$ as a function of $k_x$ and $k_z$ at the top of the VSI zone at $R = 60~\au$ in the case of $\subsc{a}{max}=10~\micron$ and $\alpha_D = 10^{-4}$. 
        In this example, we have $k_{x,\rm max} \approx 18/H_g$ and $|k_{z}|_{\rm min} \approx 1.7/H_g$ ($L \approx 0.9H_g$). 
        In the range of $k_x \leqslant k_{x,\rm max}$ and $|k_z| \geqslant |k_{z}|_{\rm min}$, $\Gamma_{\rm VSI}$ reaches a maximum of $0.003\Omega_{\rm K}$ at $k_x = k_{x,\rm max}$ and $|k_z| = |k_{z}|_{\rm min}$ (marked by the star point in Figure~\ref{fig:VSI_colormap_amax1e-3_alpha1e-4}). 
        If we did not limit $k_x$ and $|k_z|$, a higher maximum growth rate of $0.015\Omega_{\rm K}$ would be reached at the point $k_x \approx 100/H_g$ and $|k_z| \approx 1.5/H_g$ ($L \approx H_g$) marked by the filled circle in Figure~\ref{fig:VSI_colormap_amax1e-3_alpha1e-4}.
        }

        \begin{figure*}[t]
            \centering
            \includegraphics[width = 18cm,bb = 0 0 1854 615]{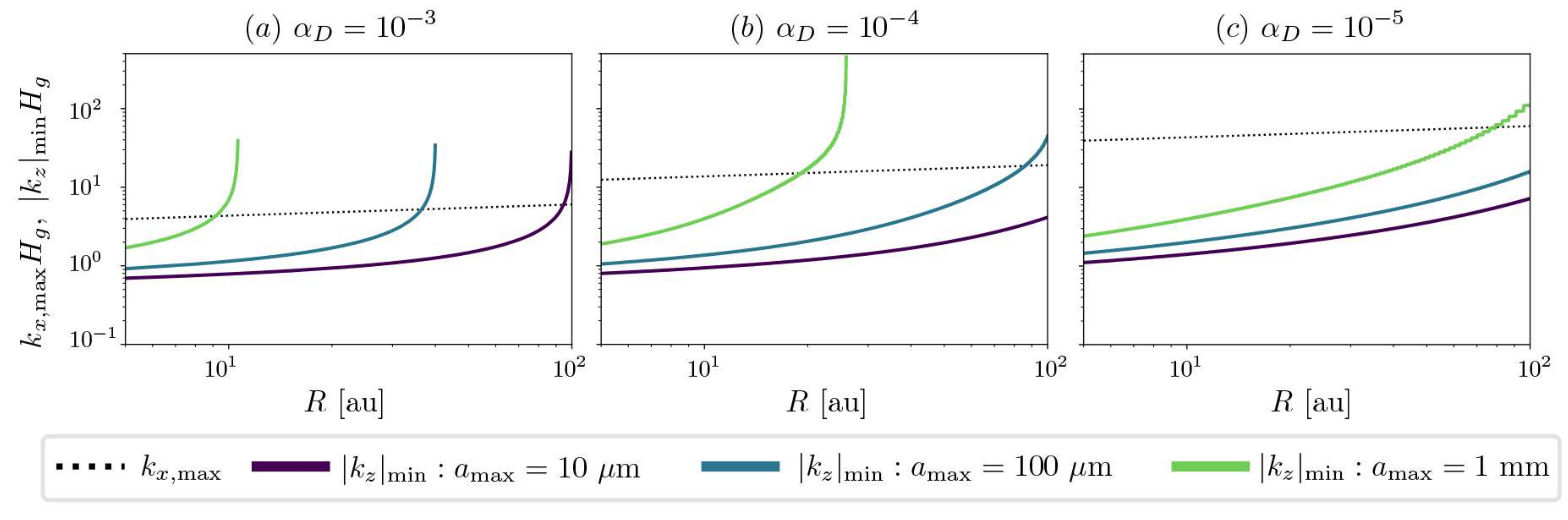}
            \caption{
            Maximum radial wavenumber $k_{x,\max}$ (dotted lines; Equation~\eqref{eq:kxmax}) and minimum vertical wavenumber $|k_z|_{\rm min}$ (solid lines; Equation~\eqref{eq:kzmin}) as a function $R$ for different values of $\alpha_D$ and $\subsc{a}{max}$. These are equal to the wavenumbers of the most unstable VSI mode (see Section~\ref{subsec:result_maximum_growth}).
            }
            \label{fig:kx_kz}
        \end{figure*}
                    
       {
       Figure~\ref{fig:kx_kz} shows the wavenumbers of the most unstable VSI modes, $k_{x,\max}$ and $|k_z|_{\rm min}$, as a function of $R$ for various values of $\alpha_D$ and $a_{\rm max}$. 
       Note that $k_{x,\max}$ depends only on $\alpha_D$ whereas $|k_z|_{\rm min}$ depends on both $\alpha_D$ and $a_{\rm max}$.
       Being inversely proportional to the VSI zone vertical thickness $L$, $|k_z|_{\rm min}$ increases as $\subsc{a}{max}$ increases or $\alpha_D$ decreases.
       Our assumption $|k_z| \ll k_x$  breaks down near the outer edge of the VSI zone where $|k_z|_{\rm min}$ diverges. However, this region is narrow compared to the VSI zone itself.
       }   

        \begin{figure*}[t]
            \centering
            \includegraphics[width = 18cm,bb = 0 0 1835 622]{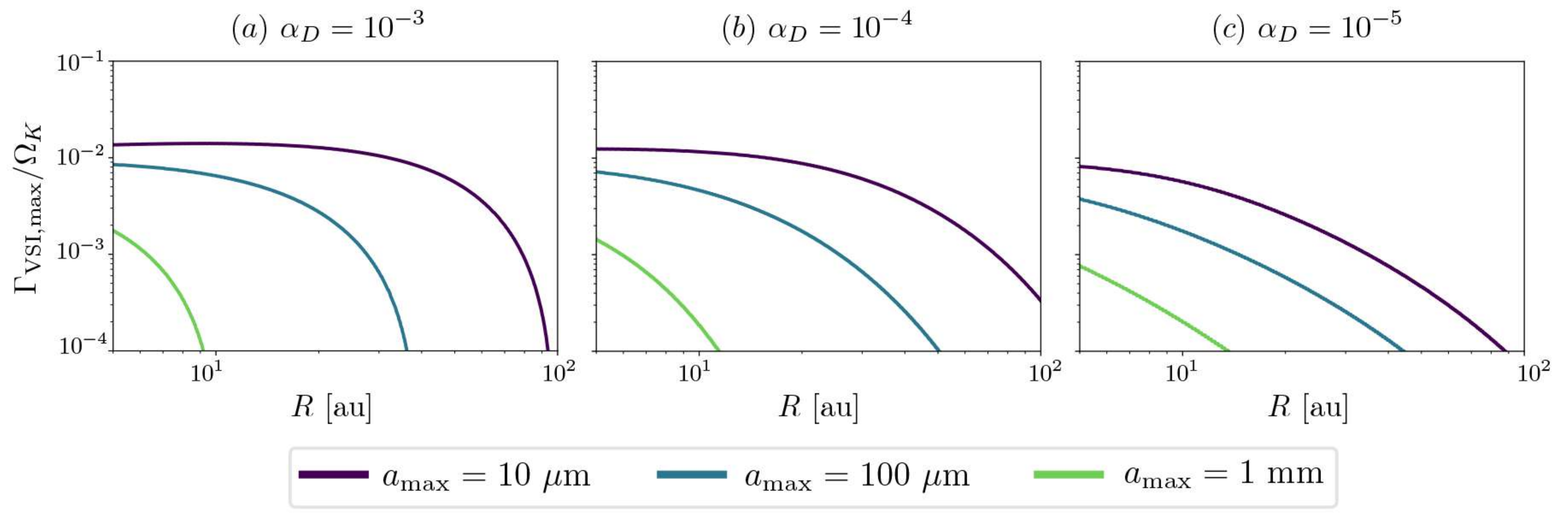}
            \caption{Maximum VSI growth rate  $\subsc{\Gamma}{VSI,max}$ as a function of $R$ for different values of $\alpha_D$ with $\subsc{a}{max} = 10~\micron$, $100~\micron$, and $1 {\rm ~mm}$.
            }
            \label{fig:VSI_growth_rate}
        \end{figure*}

        {
        Figure~\ref{fig:VSI_growth_rate} shows the maximum VSI growth rate $\Gamma_{\rm VSI,max}$ as a function of $R$ for various values of $\alpha_{\rm D}$ and $a_{\rm max}$. 
        Overall, $\Gamma_{\rm VSI,max}$ decreases with increasing $a_{\rm max}$, reflecting the fact that the VSI zone shrinks as $a_{\rm max}$ increases. 
        In this default disk model of $M_{\rm disk} = 0.01 M_\sun$, an increase in $a_{\rm max}$ from 10 $\micron$ to 100 $\micron$ causes a decrease in $\Gamma_{\rm VSI,max}$ by a factor of more than 10 at all $R \ga 5~\rm au$, and increasing $a_{\rm max}$ to 1 mm completely stabilizes the VSI at $R > $ 10 au (but see Section~\ref{subsec:result_mass} for the dependence on the disk mass).
        The maximum growth rate decreases mainly because the vertical shear $\partial(R\Omega)/\partial z$ at the top of the VSI zone, where we evaluate $\Gamma_{\rm VSI,max}$, decreases as the VSI zone shrinks vertically (see Figure~\ref{fig:NzOmegaVertical}(c)). 
        Decreasing $\alpha_{\rm D}$ also causes a decrease in $\Gamma_{\rm VSI,max}$, but this effect is minor compared to the effect of varying $a_{\rm max}$. 
        Furthermore, decreasing $\alpha_D$ broadens the range of $k_x$ to the short wavelength side (see Equation \eqref{eq:kxmax}) and promotes instability. However, this effect plays a minor role in the variation of $\Gamma_{\rm VSI,max}$ with $\alpha_D$.
        }
        
        { 
        Since our dispersion relation assumes zero viscosity and infinitesimally short cooling times, it is likely to underestimate the growth rates of modes with $k_x \sim k_{x,\rm max}$ and $|k_z| \sim |k_z|_{\rm min}$, respectively. 
        Because the most unstable modes in our analysis have both $k_x = k_{x,\rm max}$ and $|k_z| = |k_z|_{\rm min}$, the maximum growth rate of the VSI predicted from our analysis should be regarded as an upper limit. 
        }
        
    \subsection{Variation of the VSI Zone Size with the Dust-to-Gas Ratio and Disk Mass}\label{subsec:result_mass}

        \begin{figure*}[t]
                \centering
                \includegraphics[width = 18cm,bb = 0 0 1815 512]{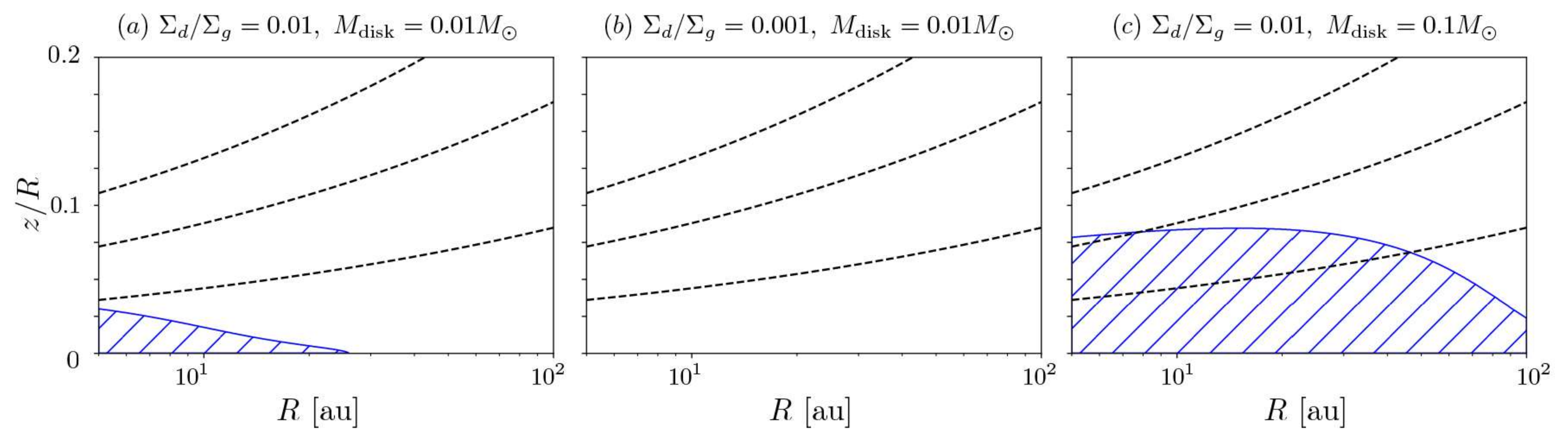}
                \caption{Location of the VSI zone (shaded area) for different values of $\Sigma_d/\Sigma_g$ and $M_{\rm disk}$ with $\subsc{a}{max}=1~{\rm mm}$ and $\alpha_D = 10^{-4}$. The dashed lines represent $H_g$, $2H_g$, and $3H_g$ in height from the midplane.}
                \label{fig:VSIzone_amount_of_dust}
        \end{figure*}
        
        { So far we have fixed the dust-to-gas mass ratio and disk mass to  $\Sigma_d/\Sigma_g = 0.01$ and $M_{\rm disk} = 0.01M_\sun$. 
        The dust-to-gas ratio can decrease with time as dust particles tend to drift toward the central star owing to disk gas drag \citep{Brauer:2008aa}. 
        The disk mass can also depend on  disk age; in particular, very young disks like the one around HL Tau can be as massive as $M_{\rm disk} \sim 0.1M_\sun$ \citep{Kwon:2015aa}.  
        As highlighted in Section \ref{subsec:result_unstable_area}, 
        the size of the VSI zone is controlled by $\subsc{\ell}{gd}$, with larger $\subsc{\ell}{gd}$ leading to smaller VSI zones.  
        Because $\ell_{\rm gd}(z=0) \propto \Sigma_{\rm g}^{1/2}/\Sigma_{\rm d} \propto (\Sigma_{\rm d}/\Sigma_{\rm g})^{-1}M_{\rm disk}^{-1/2}$ (see Equation~\eqref{eq:ldg_approx}), the VSI zone shrinks and expands as $\Sigma_{\rm d}/\Sigma_{\rm g}$ decreases and $M_{\rm disk}$ increases, respectively.
        }
        
        { We illustrate this in Figure \ref{fig:VSIzone_amount_of_dust}, where we show the maps of the VSI zone for a disk with a lower dust-to-gas ratio of  $\Sigma_d/\Sigma_g = 0.001$ and for a disk with a higher disk mass of $M_{\rm disk} = 0.1~ M_\sun$ (panels (b) and (c), respectively), both with $\subsc{a}{max}=1 {\rm ~mm}$ and $\alpha_D = 10^{-4}$.  
        We find that the VSI zone is completely removed from $R > 5~\rm au$ for $\Sigma_d/\Sigma_g  = 0.001$, whereas it extends out to 100 au for $M_{\rm disk} = 0.1M_{\odot}$ even with $a_{\rm max} = 1~\rm mm$.
        The radial extent of the VSI zone for the latter case is almost the same as that for $M_{\rm disk} = 0.01M_\sun$ and $a_{\rm max} = 10~\micron$ (see Figure~\ref{fig:VSIarea_alpha1e-4_amax}(a)), consistent with the scaling $\ell_{\rm gd}(z=0) \propto a_{\rm max}^{1/2}(\Sigma_{\rm d}/\Sigma_{\rm g})^{-1}M_{\rm disk}^{-1/2}$ from Equation~\eqref{eq:ldg_approx}. 
        }
 
\section{Discussion}\label{sec:discussion}

    \subsection{Correspondence between Vertically Local and Global Analyses}\label{subsec:discuss_appropriate}
        { 
        As mentioned in Section~\ref{sec:model}, 
        we have employed the vertically local approximation to treat the modes within a VSI zone.
        For vertically wide ($L\sim H_g$) VSI zones, our local modes should correspond to some modes in the vertically global linear analysis. 
        
        Below,  we show that the modes with the longest vertical wavelength ($|k_z| = |k_z|_{\rm min}$, Equation \eqref{eq:kzmin}) indeed correspond to the fundamental corrugation modes in the global analysis.        
        The fundamental corrugation modes are the modes with uniform vertical motion and are known to dominate the nonlinear phase of the VSI \citep{NelsonGresselUmurhan2013}.
        In the limit of short cooling times, the vertically global dispersion relation for the fundamental corrugation modes can be written as \citep{LinYoudin2015}
        \begin{equation}
        \label{eq:LY15}
            \omega^2 = \frac{1+ihq\hat{k}_x}{1+\hat{k}_x^2}\Omega_{\rm K}^2,
        \end{equation}
        where $h=H_g/R$ and $\hat{k}_x = k_xH_g$. 
        To show the correspondence between these vertically uniform modes and our local VSI modes with $|k_z| = |k_z|_{\rm min}$, we focus on local inertial modes and neglect the $\omega^4$ in our local dispersion relation (Equation \eqref{eq:bunsan2}).
        We also use $\pd v_{y0}/\pd z \sim (qH_g/R)\Omega_{\rm K}$, $N_z \ll \kappa_0 \approx \Omega_{\rm K}$, $g \sim \Omega_{\rm K}^2H_g$, and $k_z = -|k_z|_{\rm min} \sim -1/H_g$ for $L \sim H_g$. 
        Our dispersion relation then reduces to
        \begin{equation}\label{eq:omega_highL}
            \omega^2 \sim \frac{1-2\hat{k}_x|q|h+ihq\hat{k}_x}{2+\hat{k}_x^2}\Omega_{\rm K}^2,
        \end{equation}
        Since $\hat{k}_x|q|h \leq \hat{k}_{x,\rm max}|q|h \lesssim 1$ for $\alpha_D \gtrsim 10^{-5}$ (see Equation \eqref{eq:kxmax}), Equation \eqref{eq:omega_highL} agrees with Equation~\eqref{eq:LY15} to within a factor of order unity. 
        Because the most unstable modes in our analysis have $|k_z| = |k_z|_{\rm min}$, our $\Gamma_{\rm VSI,\rm max}$ serves as a good estimate for the growth rate of the fundamental corrugation modes when $L \sim H_g$. 
        }
        
    \subsection{Can VSI-driven Turbulence Stably Sustain Vertical Dust Distribution? }\label{subsec:discuss_feedback}
    { The main limitation of our model is that it has to assume the strength of disk turbulence to calculate the vertical dust distribution.
    In reality, in outer disk regions where ambipolar diffusion suppresses MRI, the VSI itself can be the dominant source of disk turbulence. 
    If this is the case, our model effectively assumes that the strength of vertical diffusion caused by  VSI-driven turbulence matches the diffusion strength $\alpha_D$ required to sustain the vertical dust distribution.
    The system would evolve until two diffusion coefficients match, but we cannot tell if such equilibrium states would exist because our current model relying on linear stability analysis does not predict the strength of VSI-driven turbulence.
    }
    
    { Moreover, even if there exists an equilibrium state, the state can be unstable against perturbations to the vertical dust distribution. 
    Unstable equilibrium is expected if, for instance, a small decrease in the the dust scale height causes a large decrease in the VSI turbulence strength, in which case dust settling would proceed in a runaway fashion .
    Dust settling also introduces effective vertical buoyancy that further stabilizes the VSI in the settled dust layers \citep{Lin:2019aa,Schafer:2020aa}.
    The stability of the system should be studied in future hydrodynamical simulations that include both the thermal and frictional coupling between the gas and dust.
    }

    \subsection{Implications for Dust Growth, Settling, and Planetesimal Formation in Outer Disk Regions}\label{subsec:discuss_formation}
        { 
        We have shown that dust growth can substantially suppresses the VSI beyond 10 au.
        This suggests that disk turbulence in this outer disk region would become weaker as dust grows.
        This may provide positive feedback to dust growth and also to planetesimal formation. Weaker turbulence would suppress collisional fragmentation of the dust particles and thereby further promote dust growth \citep[e.g.,][]{Brauer:2008aa,Birnstiel+2010,Okuzumi:2012aa}.
        Weak turbulence are also preferred for planetesimal formation via the streaming and gravitational instabilities, both of which require substantial dust settling toward the midplane \citep{Sekiya:1998aa,Youdin:2002aa,Johansen:2009aa}.
        These positive feedback effects are potentially important for understanding planet formation and dust ring/gap formation in outer regions of protoplanetary disks. 
        } 
        
        { Suppression of the VSI at large radial distances due to dust growth may also explain the high degree of dust settling in the HL Tau disk \citep{Pinte:2016aa}.
        However, quantitative estimates for the strength of VSI-driven turbulence are needed to test this hypothesis because, as we have seen in Section~\ref{subsec:result_mass}, the VSI tends to be vigorous in massive disks like the HL Tau disk ($M_{\rm disk} \sim 0.1M_\sun$; \citealt{Kwon:2015aa}). Very recently, \citet{Doi:2021aa} measured the vertical thicknesses of two dust rings at 70 and 100 au in the massive disk around HD 163296, and showed that the outer ring is much thinner than the gas disk but the inner ring is as thick as the gas disk. This may suggest that the VSI is active at $\sim 70~\rm au$ in this disk (see \citealt{Bi:2021tg} and \citealt{Binkert:2021wq} for another potential interpretation). We plan to address these open issues in future work.       
        }
  
\section{Conclusions}\label{sec:conclusions}
    {
    We have investigated the impacts of dust growth and settling on the VSI in the outer regions of protoplanetary disks using a model based on linear stability analysis.
    Our key findings are summarized as follows.
    \begin{enumerate}
        \item For fixed dust particle size distribution, a higher degree of dust settling (corresponding to a lower value of turbulence strength $\alpha_D$) leads to a VSI zone that is more confined to the midplane and more extended to larger radial distances (Figure \ref{fig:VSIarea_amax1e-2_alpha}). 
        This is because in outer disk regions of low optical depths, dust settling causes a decrease and an increase in the timescale of thermal conduction from gas to dust ($\propto \ell_{\rm gd}$;  Equation~\eqref{eq:taucoll}) at the midplane and well above the midplane, respectively (Figure~\ref{fig:Atot_amax1e-2_alpha}). 
        \item For fixed turbulence strength, the VSI zone shrinks toward the midplane and also toward the central star as dust particles grow (Figure~\ref{fig:VSIarea_alpha1e-4_amax}). Dust growth also causes dust settling, but the decrease in the total surface area of the particles due to dust growth dominates the change in the thermal relaxation timescale (Figure~\ref{fig:Atot_alpha1e-4_amax}).
        \item The maximum growth rate of the VSI decreases as dust particles grow. In our default disk model assuming $M_{\rm disk} = 0.01M_\sun$, dust growth to 1 mm in size completely stabilizes the VSI exterior to 10 au (Figure \ref{fig:VSI_growth_rate}).
        On the other hand, the VSI is more vigorous in disks with larger disk masses  (Section~\ref{subsec:result_mass}).
    \end{enumerate}
    }

    {
    Our results suggest that dust evolution, in particular dust growth, should lead to suppression of VSI-driven turbulence. 
    This effect may enable further dust coagulation and settling and may potentially promote planetesimal formation in  outer disk regions.
    The effect may also explain the high degree of dust settling observed in the dust rings around HL Tau, but testing this hypothesis requires more qualitative investigation of the interplay between dust evolution and the nonlinear development of the VSI.
    }

\acknowledgments
    { We thank Shoji Mori for the discussions that motivated this project. We also thank the referee for a helpful report that motivated us to discuss the correspondence between local and global linear analyses. This work was supported by JSPS KAKENHI Grant Numbers JP20H01948, JP20H00182, JP19K03926, JP18H05438 and JP20J01376.}

\bibliography{reference.bib}

\begin{thebibliography}{}
\expandafter\ifx\csname natexlab\endcsname\relax\def\natexlab#1{#1}\fi
\providecommand{\url}[1]{\href{#1}{#1}}
\providecommand{\dodoi}[1]{doi:~\href{http://doi.org/#1}{\nolinkurl{#1}}}
\providecommand{\doeprint}[1]{\href{http://ascl.net/#1}{\nolinkurl{http://ascl.net/#1}}}
\providecommand{\doarXiv}[1]{\href{https://arxiv.org/abs/#1}{\nolinkurl{https://arxiv.org/abs/#1}}}

\bibitem[{{Adachi} {et~al.}(1976){Adachi}, {Hayashi}, \&
  {Nakazawa}}]{Adachi:1976uv}
{Adachi}, I., {Hayashi}, C., \& {Nakazawa}, K. 1976, Progress of Theoretical
  Physics, 56, 1756, \dodoi{10.1143/PTP.56.1756}

\bibitem[{{ALMA Partnership} {et~al.}(2015){ALMA Partnership}, {Brogan},
  {P{\'e}rez}, {Hunter}, {Dent}, {Hales}, {Hills}, {Corder}, {Fomalont},
  {Vlahakis}, {Asaki}, {Barkats}, {Hirota}, {Hodge}, {Impellizzeri}, {Kneissl},
  {Liuzzo}, {Lucas}, {Marcelino}, {Matsushita}, {Nakanishi}, {Phillips},
  {Richards}, {Toledo}, {Aladro}, {Broguiere}, {Cortes}, {Cortes}, {Espada},
  {Galarza}, {Garcia-Appadoo}, {Guzman-Ramirez}, {Humphreys}, {Jung}, {Kameno},
  {Laing}, {Leon}, {Marconi}, {Mignano}, {Nikolic}, {Nyman}, {Radiszcz},
  {Remijan}, {Rod{\'o}n}, {Sawada}, {Takahashi}, {Tilanus}, {Vila Vilaro},
  {Watson}, {Wiklind}, {Akiyama}, {Chapillon}, {de Gregorio-Monsalvo}, {Di
  Francesco}, {Gueth}, {Kawamura}, {Lee}, {Nguyen Luong}, {Mangum}, {Pietu},
  {Sanhueza}, {Saigo}, {Takakuwa}, {Ubach}, {van Kempen}, {Wootten},
  {Castro-Carrizo}, {Francke}, {Gallardo}, {Garcia}, {Gonzalez}, {Hill},
  {Kaminski}, {Kurono}, {Liu}, {Lopez}, {Morales}, {Plarre}, {Schieven},
  {Testi}, {Videla}, {Villard}, {Andreani}, {Hibbard}, \&
  {Tatematsu}}]{ALMA+2014}
{ALMA Partnership}, {Brogan}, C.~L., {P{\'e}rez}, L.~M., {et~al.} 2015, \apjl,
  808, L3, \dodoi{10.1088/2041-8205/808/1/L3}

\bibitem[{{Andrews}(2020)}]{Andrews:2020aa}
{Andrews}, S.~M. 2020, \araa, 58, 483,
  \dodoi{10.1146/annurev-astro-031220-010302}

\bibitem[{{Andrews} {et~al.}(2018){Andrews}, {Huang}, {P{\'e}rez}, {Isella},
  {Dullemond}, {Kurtovic}, {Guzm{\'a}n}, {Carpenter}, {Wilner}, {Zhang}, {Zhu},
  {Birnstiel}, {Bai}, {Benisty}, {Hughes}, {{\"O}berg}, \&
  {Ricci}}]{Andrews+2018}
{Andrews}, S.~M., {Huang}, J., {P{\'e}rez}, L.~M., {et~al.} 2018, \apjl, 869,
  L41, \dodoi{10.3847/2041-8213/aaf741}

\bibitem[{{Arlt} \& {Urpin}(2004)}]{ArltUrpin2004}
{Arlt}, R., \& {Urpin}, V. 2004, \aap, 426, 755,
  \dodoi{10.1051/0004-6361:20035896}

\bibitem[{{Bai}(2015)}]{Bai2015}
{Bai}, X.-N. 2015, \apj, 798, 84, \dodoi{10.1088/0004-637X/798/2/84}

\bibitem[{{Balbus} \& {Hawley}(1991)}]{BalbusHawley1991}
{Balbus}, S.~A., \& {Hawley}, J.~F. 1991, \apj, 376, 214,
  \dodoi{10.1086/170270}

\bibitem[{{Barker} \& {Latter}(2015)}]{Barker:2015tt}
{Barker}, A.~J., \& {Latter}, H.~N. 2015, \mnras, 450, 21,
  \dodoi{10.1093/mnras/stv640}

\bibitem[{{Bi} {et~al.}(2021){Bi}, {Lin}, \& {Dong}}]{Bi:2021tg}
{Bi}, J., {Lin}, M.-K., \& {Dong}, R. 2021, arXiv e-prints, arXiv:2103.09254.
\newblock \doarXiv{2103.09254}

\bibitem[{{Binkert} {et~al.}(2021){Binkert}, {Szul{\'a}gyi}, \&
  {Birnstiel}}]{Binkert:2021wq}
{Binkert}, F., {Szul{\'a}gyi}, J., \& {Birnstiel}, T. 2021, arXiv e-prints,
  arXiv:2103.10177.
\newblock \doarXiv{2103.10177}

\bibitem[{{Birnstiel} {et~al.}(2010){Birnstiel}, {Dullemond}, \&
  {Brauer}}]{Birnstiel+2010}
{Birnstiel}, T., {Dullemond}, C.~P., \& {Brauer}, F. 2010, \aap, 513, A79,
  \dodoi{10.1051/0004-6361/200913731}

\bibitem[{{Birnstiel} {et~al.}(2011){Birnstiel}, {Ormel}, \&
  {Dullemond}}]{Birnstiel:2011aa}
{Birnstiel}, T., {Ormel}, C.~W., \& {Dullemond}, C.~P. 2011, \aap, 525, A11,
  \dodoi{10.1051/0004-6361/201015228}

\bibitem[{{Bitsch} {et~al.}(2015){Bitsch}, {Johansen}, {Lambrechts}, \&
  {Morbidelli}}]{Bitsch:2015aa}
{Bitsch}, B., {Johansen}, A., {Lambrechts}, M., \& {Morbidelli}, A. 2015, \aap,
  575, A28, \dodoi{10.1051/0004-6361/201424964}

\bibitem[{{Bohren} \& {Huffman}(1983)}]{Bohren:1983aa}
{Bohren}, C.~F., \& {Huffman}, D.~R. 1983, {Absorption and scattering of light
  by small particles}

\bibitem[{{Brauer} {et~al.}(2008){Brauer}, {Dullemond}, \&
  {Henning}}]{Brauer:2008aa}
{Brauer}, F., {Dullemond}, C.~P., \& {Henning}, T. 2008, \aap, 480, 859,
  \dodoi{10.1051/0004-6361:20077759}

\bibitem[{{Carrera} {et~al.}(2015){Carrera}, {Johansen}, \&
  {Davies}}]{Carrera:2015aa}
{Carrera}, D., {Johansen}, A., \& {Davies}, M.~B. 2015, \aap, 579, A43,
  \dodoi{10.1051/0004-6361/201425120}

\bibitem[{{Chiang} \& {Goldreich}(1997)}]{Chiang:1997aa}
{Chiang}, E.~I., \& {Goldreich}, P. 1997, \apj, 490, 368,
  \dodoi{10.1086/304869}

\bibitem[{{Chokshi} {et~al.}(1993){Chokshi}, {Tielens}, \&
  {Hollenbach}}]{Chokshi:1993aa}
{Chokshi}, A., {Tielens}, A.~G.~G.~M., \& {Hollenbach}, D. 1993, \apj, 407,
  806, \dodoi{10.1086/172562}

\bibitem[{{Cui} \& {Bai}(2020)}]{Cui:2020aa}
{Cui}, C., \& {Bai}, X.-N. 2020, \apj, 891, 30,
  \dodoi{10.3847/1538-4357/ab7194}

\bibitem[{{Doi} \& {Kataoka}(2021)}]{Doi:2021aa}
{Doi}, K., \& {Kataoka}, A. 2021, arXiv e-prints, arXiv:2102.06209.
\newblock \doarXiv{2102.06209}

\bibitem[{{Dominik} \& {Tielens}(1997)}]{Dominik:1997aa}
{Dominik}, C., \& {Tielens}, A.~G.~G.~M. 1997, \apj, 480, 647,
  \dodoi{10.1086/303996}

\bibitem[{{Dubrulle} {et~al.}(1995){Dubrulle}, {Morfill}, \&
  {Sterzik}}]{Dubrulle+1995}
{Dubrulle}, B., {Morfill}, G., \& {Sterzik}, M. 1995, \icarus, 114, 237,
  \dodoi{10.1006/icar.1995.1058}

\bibitem[{{Flaherty} {et~al.}(2020){Flaherty}, {Hughes}, {Simon}, {Qi}, {Bai},
  {Bulatek}, {Andrews}, {Wilner}, \& {K{\'o}sp{\'a}l}}]{Flaherty:2020aa}
{Flaherty}, K., {Hughes}, A.~M., {Simon}, J.~B., {et~al.} 2020, \apj, 895, 109,
  \dodoi{10.3847/1538-4357/ab8cc5}

\bibitem[{{Flaherty} {et~al.}(2015){Flaherty}, {Hughes}, {Rosenfeld},
  {Andrews}, {Chiang}, {Simon}, {Kerzner}, \& {Wilner}}]{Flaherty:2015aa}
{Flaherty}, K.~M., {Hughes}, A.~M., {Rosenfeld}, K.~A., {et~al.} 2015, \apj,
  813, 99, \dodoi{10.1088/0004-637X/813/2/99}

\bibitem[{{Flaherty} {et~al.}(2018){Flaherty}, {Hughes}, {Teague}, {Simon},
  {Andrews}, \& {Wilner}}]{Flaherty:2018aa}
{Flaherty}, K.~M., {Hughes}, A.~M., {Teague}, R., {et~al.} 2018, \apj, 856,
  117, \dodoi{10.3847/1538-4357/aab615}

\bibitem[{{Flaherty} {et~al.}(2017){Flaherty}, {Hughes}, {Rose}, {Simon}, {Qi},
  {Andrews}, {K{\'o}sp{\'a}l}, {Wilner}, {Chiang}, {Armitage}, \&
  {Bai}}]{Flaherty:2017aa}
{Flaherty}, K.~M., {Hughes}, A.~M., {Rose}, S.~C., {et~al.} 2017, \apj, 843,
  150, \dodoi{10.3847/1538-4357/aa79f9}

\bibitem[{{Flock} {et~al.}(2017){Flock}, {Nelson}, {Turner}, {Bertrang},
  {Carrasco-Gonz{\'a}lez}, {Henning}, {Lyra}, \& {Teague}}]{FlockNelson+2017}
{Flock}, M., {Nelson}, R.~P., {Turner}, N.~J., {et~al.} 2017, \apj, 850, 131,
  \dodoi{10.3847/1538-4357/aa943f}

\bibitem[{{Flock} {et~al.}(2020){Flock}, {Turner}, {Nelson}, {Lyra}, {Manger},
  \& {Klahr}}]{Flock:2020aa}
{Flock}, M., {Turner}, N.~J., {Nelson}, R.~P., {et~al.} 2020, \apj, 897, 155,
  \dodoi{10.3847/1538-4357/ab9641}

\bibitem[{{Fricke}(1968)}]{Fricke:1968aa}
{Fricke}, K. 1968, \zap, 68, 317

\bibitem[{{Goldreich} \& {Schubert}(1967)}]{GS67}
{Goldreich}, P., \& {Schubert}, G. 1967, \apj, 150, 571, \dodoi{10.1086/149360}

\bibitem[{{Goldreich} \& {Ward}(1973)}]{Goldreich:1973aa}
{Goldreich}, P., \& {Ward}, W.~R. 1973, \apj, 183, 1051, \dodoi{10.1086/152291}

\bibitem[{{Hartmann} {et~al.}(1998){Hartmann}, {Calvet}, {Gullbring}, \&
  {D'Alessio}}]{Hartmann:1998aa}
{Hartmann}, L., {Calvet}, N., {Gullbring}, E., \& {D'Alessio}, P. 1998, \apj,
  495, 385, \dodoi{10.1086/305277}

\bibitem[{{Johansen} \& {Youdin}(2007)}]{JohansenYoudin2007}
{Johansen}, A., \& {Youdin}, A. 2007, \apj, 662, 627, \dodoi{10.1086/516730}

\bibitem[{{Johansen} {et~al.}(2009){Johansen}, {Youdin}, \& {Mac
  Low}}]{Johansen:2009aa}
{Johansen}, A., {Youdin}, A., \& {Mac Low}, M.-M. 2009, \apjl, 704, L75,
  \dodoi{10.1088/0004-637X/704/2/L75}

\bibitem[{{Kataoka} {et~al.}(2013){Kataoka}, {Tanaka}, {Okuzumi}, \&
  {Wada}}]{Kataoka:2013aa}
{Kataoka}, A., {Tanaka}, H., {Okuzumi}, S., \& {Wada}, K. 2013, \aap, 557, L4,
  \dodoi{10.1051/0004-6361/201322151}

\bibitem[{{Kwon} {et~al.}(2015){Kwon}, {Looney}, {Mundy}, \&
  {Welch}}]{Kwon:2015aa}
{Kwon}, W., {Looney}, L.~W., {Mundy}, L.~G., \& {Welch}, W.~J. 2015, \apj, 808,
  102, \dodoi{10.1088/0004-637X/808/1/102}

\bibitem[{{Latter} \& {Papaloizou}(2018)}]{LatterPapaloizou2018}
{Latter}, H.~N., \& {Papaloizou}, J. 2018, \mnras, 474, 3110,
  \dodoi{10.1093/mnras/stx3031}

\bibitem[{{Lin}(2019)}]{Lin:2019aa}
{Lin}, M.-K. 2019, \mnras, 485, 5221, \dodoi{10.1093/mnras/stz701}

\bibitem[{{Lin} \& {Youdin}(2015)}]{LinYoudin2015}
{Lin}, M.-K., \& {Youdin}, A.~N. 2015, \apj, 811, 17,
  \dodoi{10.1088/0004-637X/811/1/17}

\bibitem[{{Lin} \& {Youdin}(2017)}]{Lin:2017aa}
---. 2017, \apj, 849, 129, \dodoi{10.3847/1538-4357/aa92cd}

\bibitem[{{Long} {et~al.}(2018){Long}, {Pinilla}, {Herczeg}, {Harsono},
  {Dipierro}, {Pascucci}, {Hendler}, {Tazzari}, {Ragusa}, {Salyk}, {Edwards},
  {Lodato}, {van de Plas}, {Johnstone}, {Liu}, {Boehler}, {Cabrit}, {Manara},
  {Menard}, {Mulders}, {Nisini}, {Fischer}, {Rigliaco}, {Banzatti}, {Avenhaus},
  \& {Gully-Santiago}}]{Long:2018aa}
{Long}, F., {Pinilla}, P., {Herczeg}, G.~J., {et~al.} 2018, \apj, 869, 17,
  \dodoi{10.3847/1538-4357/aae8e1}

\bibitem[{{Lynden-Bell} \& {Pringle}(1974)}]{Lynden-Bell:1974aa}
{Lynden-Bell}, D., \& {Pringle}, J.~E. 1974, \mnras, 168, 603,
  \dodoi{10.1093/mnras/168.3.603}

\bibitem[{{Lyra} \& {Umurhan}(2019)}]{LyraUmurhan2019}
{Lyra}, W., \& {Umurhan}, O.~M. 2019, \pasp, 131, 072001,
  \dodoi{10.1088/1538-3873/aaf5ff}

\bibitem[{{Malygin} {et~al.}(2017){Malygin}, {Klahr}, {Semenov}, {Henning}, \&
  {Dullemond}}]{Malygin+2017}
{Malygin}, M.~G., {Klahr}, H., {Semenov}, D., {Henning}, T., \& {Dullemond},
  C.~P. 2017, \aap, 605, A30, \dodoi{10.1051/0004-6361/201629933}

\bibitem[{{Mathis} {et~al.}(1977){Mathis}, {Rumpl}, \&
  {Nordsieck}}]{Mathis:1977aa}
{Mathis}, J.~S., {Rumpl}, W., \& {Nordsieck}, K.~H. 1977, \apj, 217, 425,
  \dodoi{10.1086/155591}

\bibitem[{{Nelson} {et~al.}(2013){Nelson}, {Gressel}, \&
  {Umurhan}}]{NelsonGresselUmurhan2013}
{Nelson}, R.~P., {Gressel}, O., \& {Umurhan}, O.~M. 2013, \mnras, 435, 2610,
  \dodoi{10.1093/mnras/stt1475}

\bibitem[{{Nelson} {et~al.}(2016){Nelson}, {Gressel}, \&
  {Umurhan}}]{Nelson+2016:erratum}
---. 2016, \mnras, 456, 239, \dodoi{10.1093/mnras/stv2440}

\bibitem[{{Okuzumi} \& {Hirose}(2012)}]{Okuzumi:2012aa}
{Okuzumi}, S., \& {Hirose}, S. 2012, \apjl, 753, L8,
  \dodoi{10.1088/2041-8205/753/1/L8}

\bibitem[{{Okuzumi} {et~al.}(2012){Okuzumi}, {Tanaka}, {Kobayashi}, \&
  {Wada}}]{Okuzumi+2012}
{Okuzumi}, S., {Tanaka}, H., {Kobayashi}, H., \& {Wada}, K. 2012, \apj, 752,
  106, \dodoi{10.1088/0004-637X/752/2/106}

\bibitem[{{Pfeil} \& {Klahr}(2019)}]{PfeilKlahr2019}
{Pfeil}, T., \& {Klahr}, H. 2019, \apj, 871, 150,
  \dodoi{10.3847/1538-4357/aaf962}

\bibitem[{{Pierens}(2021)}]{Pierens:2021ve}
{Pierens}, A. 2021, \mnras, \dodoi{10.1093/mnras/stab183}

\bibitem[{{Pinte} {et~al.}(2016){Pinte}, {Dent}, {M{\'e}nard}, {Hales}, {Hill},
  {Cortes}, \& {de Gregorio-Monsalvo}}]{Pinte:2016aa}
{Pinte}, C., {Dent}, W.~R.~F., {M{\'e}nard}, F., {et~al.} 2016, \apj, 816, 25,
  \dodoi{10.3847/0004-637X/816/1/25}

\bibitem[{{Riols} \& {Lesur}(2018)}]{Riols:2018aa}
{Riols}, A., \& {Lesur}, G. 2018, \aap, 617, A117,
  \dodoi{10.1051/0004-6361/201833212}

\bibitem[{{Sch{\"a}fer} {et~al.}(2020){Sch{\"a}fer}, {Johansen}, \&
  {Banerjee}}]{Schafer:2020aa}
{Sch{\"a}fer}, U., {Johansen}, A., \& {Banerjee}, R. 2020, \aap, 635, A190,
  \dodoi{10.1051/0004-6361/201937371}

\bibitem[{{Sekiya}(1998)}]{Sekiya:1998aa}
{Sekiya}, M. 1998, \icarus, 133, 298, \dodoi{10.1006/icar.1998.5933}

\bibitem[{{Simon} {et~al.}(2013{\natexlab{a}}){Simon}, {Bai}, {Armitage},
  {Stone}, \& {Beckwith}}]{Simon+2013a}
{Simon}, J.~B., {Bai}, X.-N., {Armitage}, P.~J., {Stone}, J.~M., \& {Beckwith},
  K. 2013{\natexlab{a}}, \apj, 775, 73, \dodoi{10.1088/0004-637X/775/1/73}

\bibitem[{{Simon} {et~al.}(2013{\natexlab{b}}){Simon}, {Bai}, {Stone},
  {Armitage}, \& {Beckwith}}]{Simon+2013b}
{Simon}, J.~B., {Bai}, X.-N., {Stone}, J.~M., {Armitage}, P.~J., \& {Beckwith},
  K. 2013{\natexlab{b}}, \apj, 764, 66, \dodoi{10.1088/0004-637X/764/1/66}

\bibitem[{{Stoll} \& {Kley}(2014)}]{StollKley2014}
{Stoll}, M. H.~R., \& {Kley}, W. 2014, \aap, 572, A77,
  \dodoi{10.1051/0004-6361/201424114}

\bibitem[{{Takahashi} \& {Inutsuka}(2014)}]{Takahashi:2014wi}
{Takahashi}, S.~Z., \& {Inutsuka}, S.-i. 2014, \apj, 794, 55,
  \dodoi{10.1088/0004-637X/794/1/55}

\bibitem[{{Takeuchi} \& {Lin}(2002)}]{TakeuchiLin2002}
{Takeuchi}, T., \& {Lin}, D.~N.~C. 2002, \apj, 581, 1344,
  \dodoi{10.1086/344437}

\bibitem[{{Tominaga} {et~al.}(2018){Tominaga}, {Inutsuka}, \&
  {Takahashi}}]{Tominaga:2018th}
{Tominaga}, R.~T., {Inutsuka}, S.-i., \& {Takahashi}, S.~Z. 2018, \pasj, 70, 3,
  \dodoi{10.1093/pasj/psx143}

\bibitem[{{Tominaga} {et~al.}(2019){Tominaga}, {Takahashi}, \&
  {Inutsuka}}]{Tominaga:2019uu}
{Tominaga}, R.~T., {Takahashi}, S.~Z., \& {Inutsuka}, S.-i. 2019, \apj, 881,
  53, \dodoi{10.3847/1538-4357/ab25ea}

\bibitem[{{Tominaga} {et~al.}(2020){Tominaga}, {Takahashi}, \&
  {Inutsuka}}]{Tominaga:2020wn}
---. 2020, \apj, 900, 182, \dodoi{10.3847/1538-4357/abad36}

\bibitem[{{Urpin}(2003)}]{Urpin2003}
{Urpin}, V. 2003, \aap, 404, 397, \dodoi{10.1051/0004-6361:20030513}

\bibitem[{{Urpin} \& {Brandenburg}(1998)}]{UrpinBrandenburg1998}
{Urpin}, V., \& {Brandenburg}, A. 1998, \mnras, 294, 399,
  \dodoi{10.1046/j.1365-8711.1998.01118.x}

\bibitem[{{van der Marel} {et~al.}(2019){van der Marel}, {Dong}, {di
  Francesco}, {Williams}, \& {Tobin}}]{van-der-Marel:2019aa}
{van der Marel}, N., {Dong}, R., {di Francesco}, J., {Williams}, J.~P., \&
  {Tobin}, J. 2019, \apj, 872, 112, \dodoi{10.3847/1538-4357/aafd31}

\bibitem[{{Weidenschilling}(1977)}]{Weidenschilling:1977wt}
{Weidenschilling}, S.~J. 1977, \mnras, 180, 57, \dodoi{10.1093/mnras/180.2.57}

\bibitem[{{Whipple}(1972)}]{Whipple:1972vv}
{Whipple}, F.~L. 1972, in From Plasma to Planet, ed. A.~{Elvius}, 211

\bibitem[{{Windmark} {et~al.}(2012){Windmark}, {Birnstiel}, {Ormel}, \&
  {Dullemond}}]{Windmark:2012aa}
{Windmark}, F., {Birnstiel}, T., {Ormel}, C.~W., \& {Dullemond}, C.~P. 2012,
  \aap, 544, L16, \dodoi{10.1051/0004-6361/201220004}

\bibitem[{{Yang} {et~al.}(2017){Yang}, {Johansen}, \& {Carrera}}]{Yang:2017aa}
{Yang}, C.~C., {Johansen}, A., \& {Carrera}, D. 2017, \aap, 606, A80,
  \dodoi{10.1051/0004-6361/201630106}

\bibitem[{{Youdin}(2011)}]{Youdin:2011aa}
{Youdin}, A.~N. 2011, \apj, 731, 99, \dodoi{10.1088/0004-637X/731/2/99}

\bibitem[{{Youdin} \& {Goodman}(2005)}]{Youdin:2005aa}
{Youdin}, A.~N., \& {Goodman}, J. 2005, \apj, 620, 459, \dodoi{10.1086/426895}

\bibitem[{{Youdin} \& {Lithwick}(2007)}]{YoudinLithwick2007}
{Youdin}, A.~N., \& {Lithwick}, Y. 2007, \icarus, 192, 588,
  \dodoi{10.1016/j.icarus.2007.07.012}

\bibitem[{{Youdin} \& {Shu}(2002)}]{Youdin:2002aa}
{Youdin}, A.~N., \& {Shu}, F.~H. 2002, \apj, 580, 494, \dodoi{10.1086/343109}

\end{thebibliography}

\appendix

\section{Thermal Relaxation Timescale}\label{appendix:timescale}
    
    \begin{figure*}
        \includegraphics[width=18cm,bb = 0 0 1525 512]{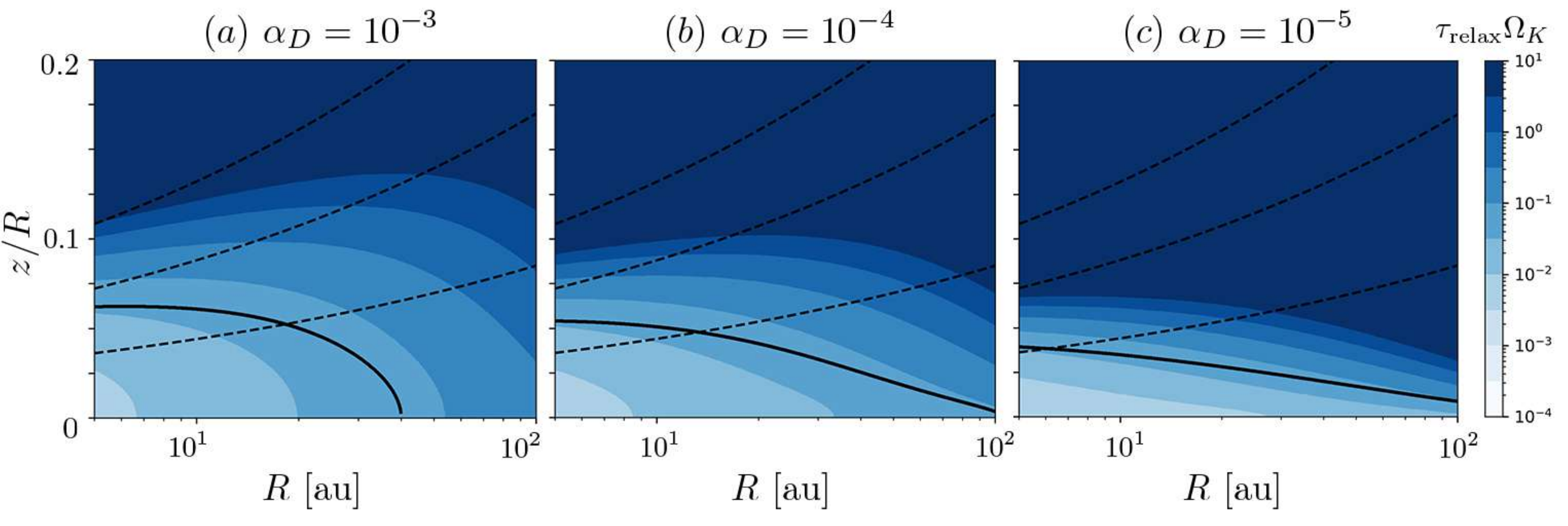}
        \caption{
        Thermal relaxation timescale $\subsc{\tau}{relax}$ normalized by $\Omega_{\rm K}^{-1}$ as a function of $R$ and $z/R$ for different values of $\alpha_D$ with $a_{\rm max}=100 {\rm ~\micron}$. The dashed lines represent $z = H_g$, $2H_g$, and $3H_g$. The solid line marks $\subsc{\tau}{relax}=\subsc{\tau}{crit}$.
        }
        \label{fig:coll_alpha}
    \end{figure*}
    \begin{figure*}
        \includegraphics[width=18cm,bb = 0 0 1525 512]{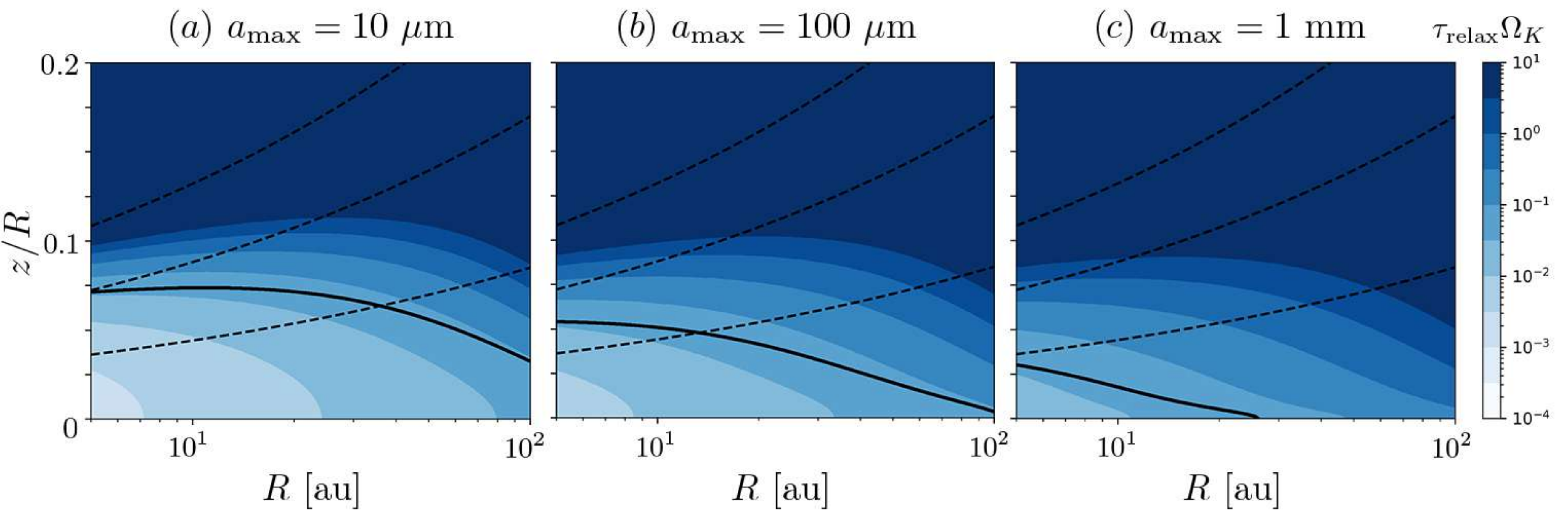}
        \caption{
        Thermal relaxation timescale $\subsc{\tau}{relax}$  normalized by $\Omega_{\rm K}^{-1}$ as a function of $R$ and $z/R$ for different values of $a_{\rm max}$ with $\alpha_D=10^{-4}$. The dashed lines represent $z = H_g$, $2H_g$, and $3H_g$. The solid line marks $\subsc{\tau}{relax}=\subsc{\tau}{crit}$.
        }
        \label{fig:coll_amax}
    \end{figure*}
    
    { Figures \ref{fig:coll_alpha} and Figure \ref{fig:coll_amax} show $\tau_{\rm relax}$ ($=\tau_{\rm coll}$) normalized by $\Omega_{\rm K}^{-1}$ as a function of $R$ and $z$ for all parameter sets considered in this study. 
    A smaller $\alpha_D$ produces a larger $\tau_{\rm relax}$ well above the midplane (because $\tau_{\rm relax} = \tau_{\rm coll} \propto \ell_{\rm gd}$; see Figure \ref{fig:Atot_amax1e-2_alpha}), resulting in a VSI zone that is more confined to the midplane (see the boundaries of the VSI zones marked by the solid lines in Figure~\ref{fig:coll_alpha}).
    A larger $a_{\rm max}$ causes an increase in $\tau_{\rm relax}$ at all heights and hence leads to a narrower VSI zone. 
    }

\end{document}